\documentclass[preprint,12pt]{elsarticle}
\usepackage{subcaption}
\usepackage{graphicx}
\usepackage{amssymb}
\usepackage{todonotes}
\usepackage{booktabs}
\usepackage{multirow}
\usepackage[hyphens]{url}

\journal{Computers and Security}
\begin{document}

\begin{frontmatter}

\title{Encrypted and Covert DNS Queries for Botnets: Challenges and Countermeasures}

\author[unipi]{Constantinos Patsakis\corref{cor1}}
\ead{kpatsak@unipi.gr}
\author[unipi]{Fran Casino}
\ead{francasino@unipi.gr}
\address[unipi]{Department of Informatics, University Piraeus, 80 Karaoli \& Dimitriou str, 18534 Piraeus, Greece}
\author[bu]{Vasilios Katos}
\ead{vkatos@bournemouth.ac.uk}
\address[bu]{Bournemouth University, Poole House P323, Talbot Campus, Fern Barrow, Poole, Dorset, BH12 5BB, UK}

\begin{abstract}
There is a continuous increase in the sophistication that modern malware exercise in order to bypass the deployed security mechanisms. A typical approach to evade the identification and potential take down of a botnet command and control server is domain fluxing through the use of Domain Generation Algorithms (DGAs). These algorithms produce a vast amount of domain names that the infected device tries to communicate with to find the C\&C server, yet only a small fragment of them is actually registered. This allows the botmaster to pivot the control and make the work of seizing the botnet control rather difficult.

Current state of the art and practice considers that the DNS queries performed by a compromised device are transparent to the network administrator and therefore can be monitored, analysed, and blocked. In this work, we showcase that the latter is a strong assumption as malware could efficiently hide its DNS queries using covert and/or encrypted channels bypassing the detection mechanisms. To this end, we discuss possible mitigation measures based on traffic analysis to address the new challenges that arise from this approach.
\end{abstract}


\end{frontmatter}

\section{Introduction}
\label{sec:intro}

The amount and sophistication of modern malware are continuously increasing creating a new industry, cybercrime, whose economy is estimated to have a capitalisation of 1.5 trillion dollars \cite{isma2018}. While this industry has many monetisation sources \cite{moore2009economics} such as spamming \cite{rao2012economics}, ad injection \cite{CHEN2017164}, denial of service and phishing to name a few, one of the most crucial aspects is the management of compromised hosts. The critical nature of the latter lies in the fact that this management should allow the adversary to (a) orchestrate further attacks by, e.g. sending his compromised hosts (bots) new commands, (b) prolong the discovery of the attack, and (c) prevent law enforcement agencies from discovering his true identity.

\begin{figure}[th]
    \centering
    \includegraphics[width=\textwidth]{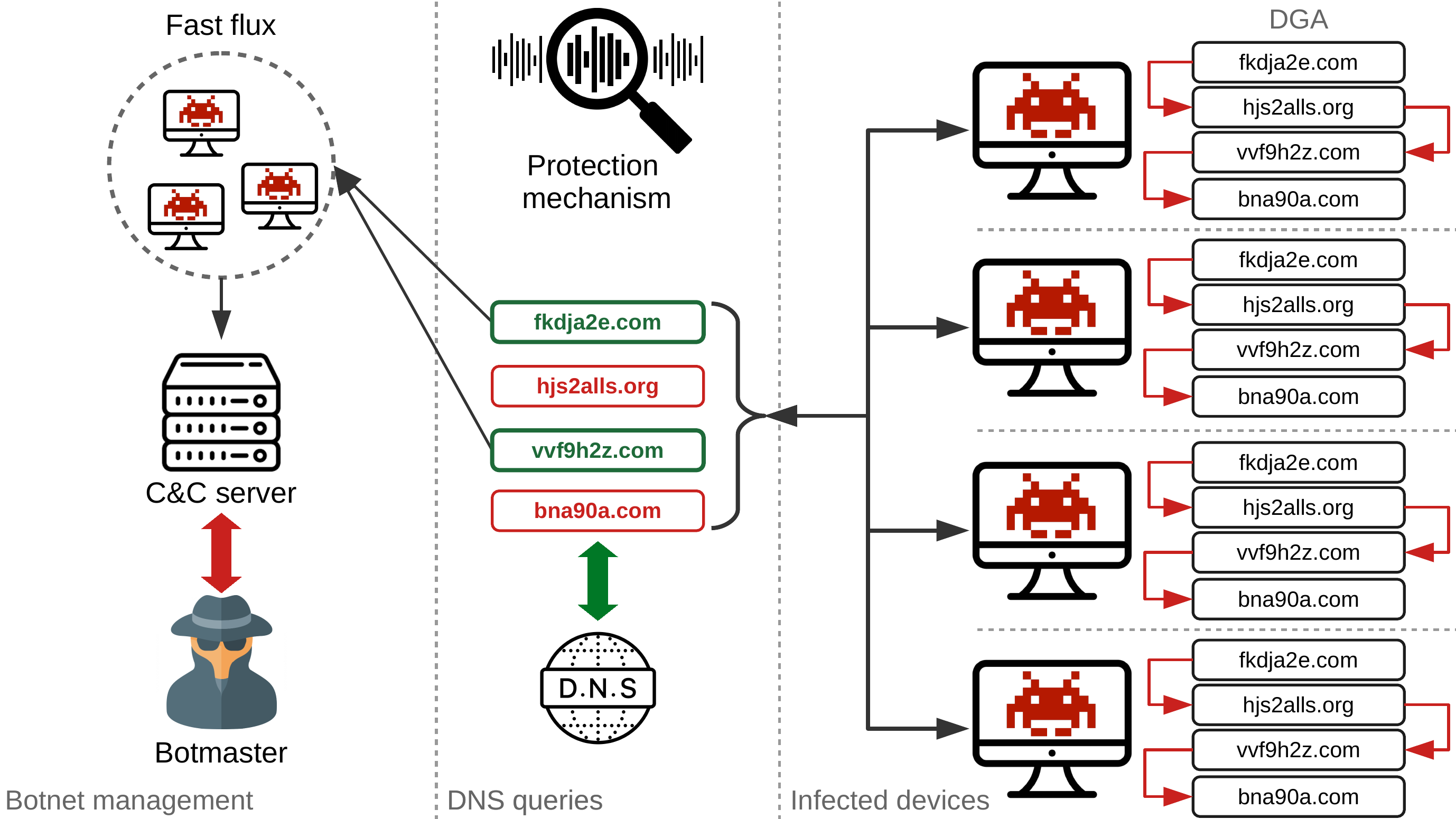}
    \caption{Botnet management with DGAs. Source \cite{ijisipfs}.}
    \label{fig:botnet_DGA}
\end{figure}

Clearly, a direct communication channel between infected devices and the Command and Control (C\&C) server can be efficiently detected and blocked  once one detects that a machine has been compromised, by blacklisting a specific IP or domain. To prevent this, malware authors try to use communication channels that disguise the traffic as benign and cannot be easily blocked, e.g. social networks or frequently change the domain names that host the C\&C server. In the latter case, which is the focus of this work, the adversary uses a Domain Generation Algorithm (DGA) that generates millions of pseudo-random domain names and the compromised devices try to connect and retrieve new commands, rendering blacklisting methods useless. However, the adversary only registers a handful of them; therefore, they can regularly pivot both the domain as well as the IP of the C\&C server without losing the control of the compromised devices. The basic model is illustrated in Figure \ref{fig:botnet_DGA} as there can be several variations. The botmaster has a deterministic pseudo-random generator (PRNG) to create a set of domain names installed in all compromised devices. As a result, these devices would periodically try to resolve these generated domain names. However, the botmaster has registered only a few of them, therefore, only those resolve to an actual machine. To further perplex possible takedown mechanisms, the botmaster uses fast flux to change the IPs that the registered domains resolve to, which may be some of the compromised devices.

\subsection{Motivation}
By design, the widely used DNS protocol is unencrypted and does not allow for authentication, allowing an adversary to initiate a wide range of attacks. The above has led to the introduction of several protocols which offer confidentiality and authentication of DNS queries and responses including among others DNSCurve, DNSCrypt and the different flavours of DNS over HTTPS/TLS/DTLS. These protocols are actively being deployed by major DNS servers; including Google, Cloudflare, and OpenDNS, while Mozilla is considering making DNS over HTTPS the default option for the DNS queries of the browser\footnote{\url{https://blog.mozilla.org/security/2019/04/09/dns-over-https-policy-requirements-for-resolvers/}}. While this may allow for increased privacy of individuals, one has to acknowledge that the fact that DNS queries are unencrypted facilitates both the detection and identification of botnets from the queries that bots make to the C\&C server. Therefore, we need to study what happens if a malware decides to use the privacy-preserving DNS services to resolve its C\&C server and whether such actions can be detected, as, by definition, current methods would be rendered useless.

\subsection{Main contributions}
The contribution of this work is twofold. First, we illustrate a critical gap in current state-of-the-art and practice methods for detecting botnets that use DGAs. More precisely, the core assumption that researchers and practitioners set to date is that they consider that an infected machine would try to connect to the C\&C server using the generated domain names from the DGA and they would be able to intercept these DNS requests. In this regard, a high rate of failed DNS queries from a specific machine implies the presence of an infected machine. Moreover, studying these requests from the perspective of a network administrator one could try to determine whether the request originates from a DGA by, e.g. evaluating the entropy of the domain name. While the hypothesis seems rather straightforward, it is rather strong as they assume that the adversary would use a plaintext channel to resolve the domain name.

In light of recent trends in malware and the continuous use of encryption \cite{hedge19}, this work investigates the challenges that arise when the adversary uses encrypted channels to perform DNS queries. While this has not been discussed in the literature nor reported by CERTs, we consider necessary to proactively discuss and study the feasibility of such an approach. Therefore, we study the possibility of botnets using protocols like DNSCurve and DNSCrypt, and mechanisms such as DNS over HTTPS/TLS to communicate with their C\&C server using connections with whitelisted domains over standard whitelisted protocols. Moreover, we investigate possible detection mechanisms of such queries and experimentally show that traffic analysis on the exchanged packets can lead to very efficient detection, in terms of both computational overhead and accuracy, for specific DGAs. Therefore, even though DNS queries can be performed over covert channels indicators of compromise (IoCs) can be composed. In fact, we illustrate that using the Hodrick-Prescott filter \cite{hodrick1997postwar} one can accurately classify them using a small amount of samples.

\subsection{Organisation of this work}
The rest of this work is structured as follows. In Section \ref{sec:back} we present the related work regarding malware and DGAs. Then, in Section \ref{sec:method} we discuss the usage of covert DNS queries from botnets using DGAs and the challenges that arise for current state of the art mechanisms. In Section \ref{sec:experiments} we present our experimental setup and the datasets we utilise. Afterwards, in Section \ref{sec:discussion} we discuss the findings of our extensive experiments using covert DNS queries and the emerging patterns that emerge and showcasing that traffic analysis can efficiently reveal such actions for specific DGAs. Finally, the article concludes discussing possible countermeasures and summarising our contributions.

\section{Related work}
\label{sec:back}

Nowadays, a common practice in malware-based campaigns is to use remote servers (i.e. C\&C servers), which send instructions/orders to infected devices \cite{203628,nadji2017still} to perform a DDoS attack for example \cite{kolias2017ddos}. The typical mechanism used in the past was to hardcode the IP addresses or possible domains of the C\&C servers in the malware, however, from the attacker's perspective, this entailed a set of drawbacks such as losing control of the botnet once the IP or domain of the C\&C server was identified. Therefore, several methods to partially or fully decentralise the management of infected devices have been implemented. With peer-to-peer botnets \cite{grizzard2007peer}, there is no centralised C\&C server, but infected hosts act as both client and servers and efficiently pass commands to each infected device. The \textit{Fast Flux} approach imitates Content Distribution Networks (CDNs) by resolving a domain name to multiple IP addresses. This can be achieved by using low TTL which forces DNS to refresh their cache associated with these domains repeatedly. To this end, a subset of the infected nodes may be used as what is called a \textit{flux agent}. A flux agent is a device whose IP is temporarily registered as the host of a domain that the bots are querying, yet they are forwarding the traffic to the C\&C server. For more about fast flux networks, the interested reader may refer to \cite{holz2008measuring} and \cite{akamai}.

To provide another layer of protection, malware use what we call \textit{Domain Name Generator Algorithms (DGAs)}. To this end, a malware uses a PRNG to create a set of domain names and this approach has become the default methodology \cite{7535098,6175908}. Hence, infected devices check the list of generated domains until they find the C\&C server, whose location may also change dynamically. In this regard, blacklisting domains is rendered useless as it implies many practical issues.

In general, DGA detection methods use simple domain name classification according to some features such as entropy, length or lexical characteristics to determine whether a DGA has generated a domain name or not, such as in \cite{Aviv2011} and \cite{6151233}.
Nevertheless, several classification approaches use additional information such as WHOIS or DNS traffic analysis to detect abnormal behaviours like a high volume of NXDomain responses, which may indicate that a device has been infected \cite{Zhou2013DGABasedBD,5762763,1}. Another technique was proposed by Yadav et al. in \cite{yadav2012}, which groups DNS queries originated by a specific client to compute the correlation between distinct requests pointing to the same domains. In a more recent work, Manadhata et al. \cite{yadavgraph} model the detection problem as a graph inference problem and construct a host-domain graph from proxy logs to classify domains into benign and malicious with a certain probability. Gong et al. \cite{gongodyseey}, perform domain name classification using a clustering algorithm with time-based information as well as IP-domain auxiliary data. Other machine learning approaches can be found in \cite{7163279} and \cite{5762763}, in which authors use a set of features extracted from network traffic and other well-known domain sources such as Alexa\footnote{https://www.alexa.com/topsites} to enhance DGA detection.



More recently, since domain names generated by DGAs exhibited a high level of randomness facilitating thus their detection, attackers adopted the use of English wordlists, which made it more difficult to discern between valid and DGA-generated domains. Similarly, the concept of \textit{domain shadowing} is also gaining attention, which relies on the use of valid domains that were previously hacked \cite{Liu2017}. Other DGAs generate domain names which have high chances of collision with valid domains and/or with other DGA malware \cite{johannesbader} so that finding and analysing them becomes more challenging.

There are some techniques to overcome such new generation of DGA algorithms such as the work proposed in \cite{18}, where authors use a short-term memory network (LSTM) to perform binary domain classification using raw domain names as features. Similar classification approaches based in neural networks can be found in \cite{attardi2018bidirectional,choudhary2018algorithmically}. In \cite{Anderson2016}, authors use a generative adversarial network (GAN) to implement a deep learning based DGA which is capable of bypass classical deep learning detectors. Subsequently, such information is used as input feedback to the DGA detectors to enhance their accuracy and robustness. In the case of \cite{stefanotracking}, the authors are able to characterize and classify similar DGA-generated domains, generating knowledge about the evolving behaviour of botnets. More recently, Curtin et al. developed the smashword score \cite{curtin2018detecting}, a new metric that uses n-gram overlapping combined with information provided from WHOIS lookups to determine whether a DGA has generated a domain name or not. For a detailed overview and classification of methods of how malicious domains can be detected, the interested reader may
refer to \cite{zhauniarovich2018survey}. Moreover, a summary of the main literature families is provided in Table \ref{tab:soasummary}. Note that none of the existing methods can analyze encrypted communications. Although some approaches do not use side information (i.e. since they only analyze domain name lexical features) as in \cite{attardi2018bidirectional,18,choudhary2018algorithmically}, the fact that DNS query responses are covert hinders the detection since we cannot check whether it was a true or false positive in real scenarios, making the discovery of novel DGA families even more difficult.

 \begin{table}[h]
   \setlength{\tabcolsep}{6pt}
   \scriptsize
   \caption{Summary of literature families and their main characteristics. NN stands for Neural Networks and RF for random-forest classifiers.}
  \begin{tabular}{cp{.17\columnwidth}p{.2\columnwidth}p{.2\columnwidth}p{.1\columnwidth}}
    \toprule
   \textbf{DGA Family}  & \textbf{Description} & \textbf{Detection features} & \textbf{Main detection methods}  &  \textbf{Covert DNS detection} \\
   \midrule
    Arithmetic-based & A PRNG generates alphanumerical combinations  & Randomness, lexical structure & entropy, lexicographic analysis, DNS response, WHOIS&  No \\
     Hash-based& Create an hex representation of a hash & Randomness, lexical structure  &  entropy, lexicographic analysis, DNS response, WHOIS & No \\
     Wordlist-based & Combination of words extracted from a dictionary & Lexical structure and n-gram analysis & Classifiers (NN, RF) &No \\
     Permutation-based & Permutations over valid or word-based domains & Lexical structure, n-gram analysis, DNS response, pattern analysis & Classifiers (NN, RF) & No \\
     \bottomrule
   \end{tabular}
   \label{tab:soasummary}
 \end{table}

With \textit{DNS tunnelling}, the malware tries to exploit the features of DNS protocol since DNS traffic is allowed unrestricted by most firewall installation. To this end, data are encoded as DNS queries and responses to avoid detection and bypass the installed security measures. Two well-known examples are Feederbot \cite{dietrich2011botnets} and Morto\footnote{\url{https://www.symantec.com/connect/blogs/morto-worm-sets-dns-record}} which used the TXT resource record to deliver their encrypted commands to the infected hosts. Nevertheless, DNS is also exploited by botnets to amplify the bandwidth of their attacks \cite{anagnostopoulos2013dns,anagnostopoulos2016new}.

It is apparent that in the literature researchers always assume that the DNS query information is plaintext or that they can collect it and analyse it. We consider that the latter is a strong assumption as DNS traffic can be encrypted or tunnelled to hide its existence. However, as recently illustrated \cite{ijisipfs}, DGAs can be extended to Resource Identifier Generation Algorithms (RIGA) which allows the use of other protocols beyond DNS and allow the orchestration to exploit other more decentralised protocols. 

\section{Covert DNS queries}
\label{sec:method}
In what follows we introduce our threat model and detail how an adversary would try to armour a malware.
\subsection{Threat model}
In our model, we assume that an adversary has managed to compromise a device in the internal network of an organisation. The adversary wants to issue commands to the compromised host (bot) without using a direct communication channel with him; therefore, she opts for the use of a DGA. Furthermore, the adversary wants to hide the existence of this mechanism by encrypting all the DNS queries and tunnel them through another protocol that is allowed from the firewall policy of the host network. To this end, we assume that the adversary has full control of the compromised host and of some other hosts outside the host network that she wants to contact. Therefore, we assume that the adversary cannot drop packets of other hosts nor change the network policies of the host network. Moreover, we assume that any unencrypted traffic will be collected by the host network and the use of any unauthorised protocol or connection with unauthorised host; both internal and external would be blocked.

\subsection{Armouring a malware}

As discussed, many malware use DGAs in their attempt to hide the actual C\&C server. However, their continuous DNS queries may trigger alerts to installed security mechanisms as they will detect a high amount of failed DNS queries to usually random looking domains from specific hosts. This is due to the very nature of the DNS protocol which does not provide data security. Apparently, the lack of these mechanisms exposes users to many threats such as eavesdropping and a set of man-in-the-middle attacks (e.g. DNS data manipulation). Nevertheless, in this scenario, it allows network administrators to monitor the domains that their hosts are trying to connect to.

While there are several solutions to counter such threats, with the most dominant one being DNSSEC \cite{ateniese2001new}, we are only interested in protocols that offer confidentiality. The reason behind this choice is that in order to armour the malware to bypass many network security mechanisms we aim to encrypt the DNS queries and/or tunnel them via other whitelisted protocols. The concept is that if the DNS queries and their responses are encrypted, then no one else could determine which are the requested domains and whether they exist. In this regard, we limit our scope to specific protocols and schemes which are deployed by legitimate DNS servers and offer confidentiality when performing DNS queries. The list is rather limited and consists of the following options:

\begin{enumerate}
\item \textbf{DNSCurve:} This protocol was proposed by Bernstein \cite{bernstein2009dnscurve} and addresses the confidentiality, integrity and availability of DNS. The protocol uses strong yet very fast encryption, however, despite its efficacy it has not been widely deployed, with the most widely used DNS server supporting it being OpenDNS\footnote{\url{https://umbrella.cisco.com/blog/2010/02/23/opendns-dnscurve/}}.
\item \textbf{DNSCrypt:} Denis and Fu introduced this protocol to authenticate communication between a DNS client and a DNS resolver \cite{denis2015dnscrypt}. Currently, it is the most widely used encrypted DNS protocol as many well-known DNS servers provide it, including among others OpenDNS and Yandex.
\item \textbf{DNS over HTTPS:} This protocol offers DNS resolution over an encrypted HTTPS connection to provide end-to-end authenticated DNS lookups. Well-known DNS providers such as Google and CleanBrowsing are already compliant with this protocol.  
\item \textbf{DNS over TLS:} As the name suggests, this protocol runs DNS transactions over TLS (see IETF RFCs 7858\footnote{https://tools.ietf.org/html/rfc7858} and 8310\footnote{https://tools.ietf.org/html/rfc8310}). Therefore, DNS queries sent to the resolver are performed in an encrypted channel, enhancing security and privacy. This solution is implemented by some DNS providers such as Cloudflare and CleanBrowsing.
\item \textbf{DNS over Datagram Transport Layer Security (DTLS):} This is an experimental protocol to offer confidentiality for DNS queries \footnote{\url{http://www.rfc-editor.org/info/rfc8094}} using UDP to improve performance and addresses packet loss and reordering that may happen during packet delivery which are traditional problems of DTLS.
\end{enumerate}

Evidently, the first two options imply some constraints in their adoption from malware as they imply the usage of ``exotic'' protocols that in, e.g. corporate/monitored environments would be blocked from the network firewall. It should be noticed that similar to HTTPS, DNSCrypt also operates in port 443. While the two protocols differ, if the firewall does not correctly identify the protocol, DNSCrypt could tunnel DNS queries and responses. However, in the case of the latter two methods, the DNS queries and responses are masqueraded under the veil of the widely used HTTPS/TLS protocols with legitimate domains. Therefore, the DNS queries can be performed in an entirely covert way as the traffic does not trigger any alert in the security mechanisms, but it is tunnelled as normal HTTPS/TLS traffic. DNS over DTLS transfers datagrams over TLS, using UDP and port 853. The protocol is quite efficient, however, it is in experimental stage and not actively used by any major provider to be evaluated.

While the use of DNSCurve and DNSCrypt is subject to firewall policy constraints, in all the cases above the host manages to efficiently resolve the IP of a domain name using strong encryption primitives. Practically, the network administrator will have to analyse packets with encrypted traffic. The interested reader may refer to \cite{anagnostopoulos2012dnssec} for a more detailed comparison of DNSSEC and DNSCurve.

\begin{figure}[th]
    \centering
    \includegraphics[width=\textwidth]{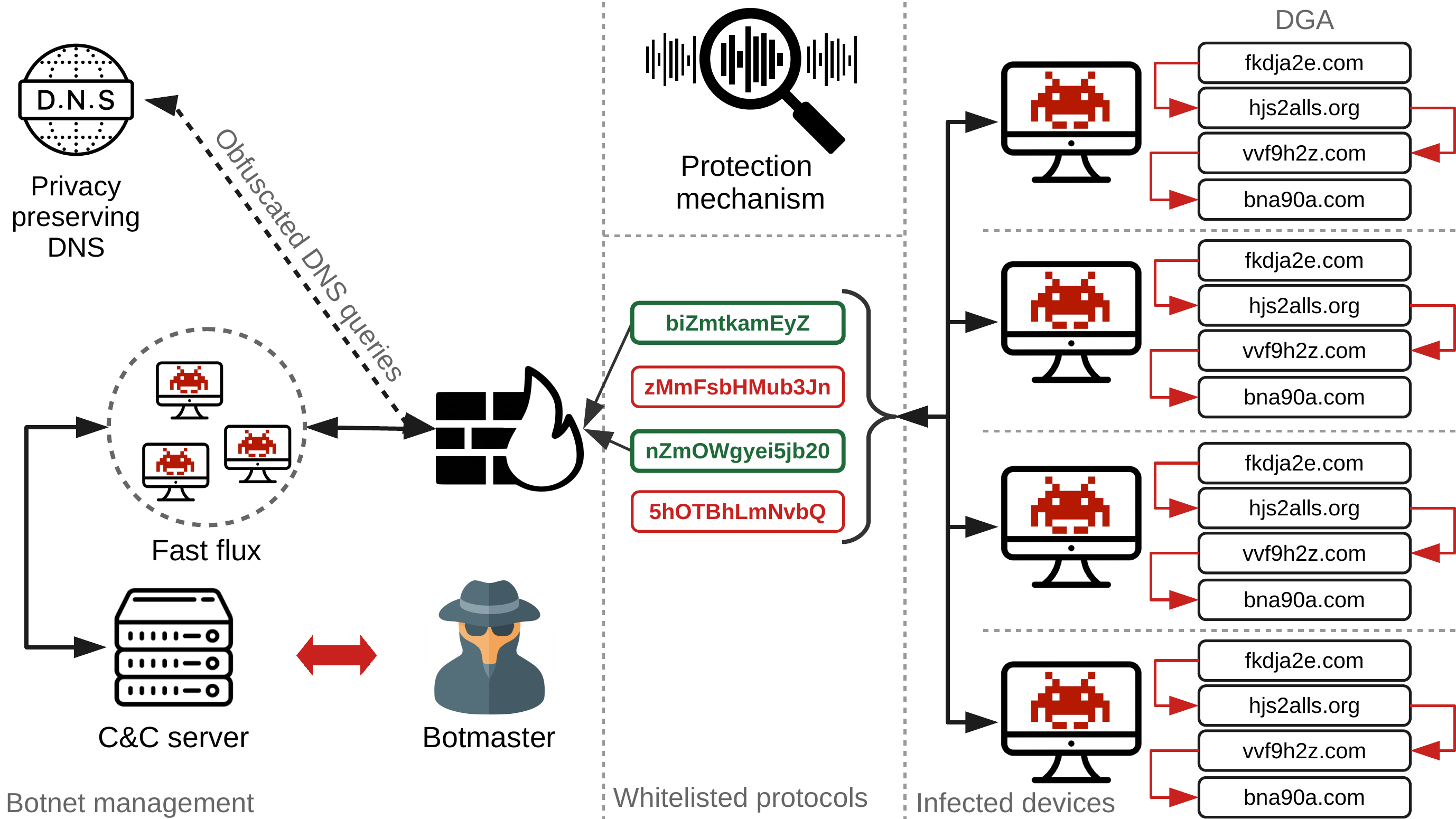}
    \caption{The armored malware.}
    \label{fig:botnet_DGA_proposed}
\end{figure}

\section{Experimental setup}
\label{sec:experiments}
The experimental evaluation aims to explore whether is it possible to construct good quality Indicators of Compromise 
to determine whether encrypted network traffic corresponds to the one generated by a botnet that using a covert channel for requesting the possible domain name of its C\&C server, as analysed in the previous section. The quality of an IoC in this paper adopts Bianco's pyramid of pain approach \cite{bianco2014}. That is, we aim for creating higher IoCs on the pain scale, such as Network Artefacts or Tools, as the lower IoCs cannot be considered reliable due to the encryption layer.
To this end, we assume that the network administrator would perform traffic analysis on the packets on the intercepted HTTPS/TLS traffic. In our experiments, we do not consider DNSCrypt and DNSCurve as the protocols can be easily detected and blocked by a firewall. 
The reader should note that by following this approach, the adversary manages to have the security and privacy guarantees of standard TLS for her queries. Therefore, it is practically impossible to perform any attack unless there are implementation issues, at the trade-off of linear complexity to the length of exchanged information for using standard TLS communication.

To investigate the capabilities and approaches of traffic analysis in the present problem domain, synthetic datasets were constructed. These datasets consist of covert DNS queries to registered and non-registered domains. For the former, we have used the Alexa top 1000 domains, while for the latter we use non registered domain names generated from ten different DGAs. More precisely, we have used the DGAs presented in Table \ref{tbl:dataset_properties} which have been published by Abakumov \footnote{\url{https://github.com/andrewaeva/DGA}}. In columns Min, Max, Average and Stdev we refer to the corresponding measures for the domain length of each dataset. Unique values indicate whether all possible lengths of domain names are represented in each dataset. For instance, in Conflicker all the possible values (16-8+1=9) are represented in the dataset, however, this is not the case for, e.g. Alexa as we have 22 unique values in the dataset and there could be 25 unique values, showing that domains of 3 specific lengths do not exist in the dataset.

From each DGA we have kept the first 1000 domains. This procedure creates a rich dataset of 11.000 domains of existing and non-existing domains. Finally, we collect the generated traffic and try to correlate the results to determine whether a device performs covert DNS queries. 

\begin{table}[th!]
    \centering
    \begin{tabular}{lrrrrr}\toprule
    \textbf{Dataset} & \textbf{Min} & \textbf{Max} & \textbf{Average} & \textbf{Stdev} & \textbf{Unique values} \\\midrule
        Alexa & 4 & 28 & 11.349 & 3.237 & 22\\
         Conflicker & 8 & 16 & 11.755 & 1.983 & 9\\
CryptoLocker & 15 & 21 & 17.783 & 1.424 & 7\\
GOZ & 20 & 35 & 28.241 & 2.431 & 16\\
Matsnu & 28 & 40 & 30.527 & 2.038 & 13\\
new GOZ & 26 & 32 & 29.885 & 1.087 & 7\\
Pushdo & 11 & 11 & 11.000 & 0.000 & 1\\
Ramdo & 20 & 20 & 20.000 & 0.000 & 1\\
Rovnix & 24 & 38 & 26.794 & 2.622 & 15\\
Tinba & 16 & 16 & 16.000 & 0.000 & 1\\
Zeus & 26 & 32 & 29.878 & 1.038 & 7\\\bottomrule
    \end{tabular}
    \caption{Datasets used in the experiments and their statistical properties in terms of domain length characteristics.}
    \label{tbl:dataset_properties}
\end{table}

Our methodology for synthesizing our datasets is rather straightforward. For each of the datasets mentioned above, we perform DNS queries over HTTPS and TLS and intercept the generated traffic using the well-known \texttt{tcpdump} tool. To facilitate the process and to allow for further validation of the results, we use \texttt{pydig}\footnote{\url{https://github.com/shuque/pydig}} to perform all the covert DNS queries. Then, we parse the generated \texttt{pcap} file and perform feature extraction on each captured packet. Using \texttt{tshark}; the command line version of \texttt{Wireshark}, and Linux command line tools, we filter the traffic to extract the necessary packages from the \texttt{pcap} files. The features that we extract from each packet are the source IP, the target IP, the size of each packet, the protocol and the info that \texttt{tshark} produces. The latter are some metadata for each package that facilitate the filtering process. Based on these features we investigate whether they can be used independently and in combination to determine whether a machine performs covert DNS queries. To allow the testing and validation of our claims and methodology we have uploaded all the needed scripts and sample datasets in GitHub\footnote{\url{https://github.com/kpatsakis/covert_dns_queries}} in the form of \texttt{pcap} files.

The aforementioned steps create 22 datasets: 11 for DNS over TLS and 11 for DNS over HTTPS. We argue that the packages of the host should not be considered as their content may greatly vary depending on the client-side implementation. Therefore we limit our results to the packages received from the DNS server over HTTPS/TLS. From these packages, we omit the packages which are the initiation or the termination of TLS as they do not contain actual information for the DNS query but other information such as parameter and cipher negotiation. Therefore, the rest of the packages must contain the actual response for the query to the DNS server. As such, the latter coincides with the experimental results.

The basic concept behind this choice of features is that the response of the DNS server in terms of length for NXdomains will differ significantly from the response for existing domains. In the latter case, the response should contain information like the IP address etc. making it longer in principle. The specifications of DNS in RFC 1035 \cite{mockapetris2004rfc} for responses to DNS queries justify this intuitive result. To facilitate the reader, we provide the output of some tools, for instance, see Figure \ref{fig:dig_Google} and \ref{fig:dig_DGA} which illustrate the output of two DNS queries using \texttt{dig}. Clearly, the response of an existing domain contains ``different'' information in terms of both quality and quantity. Therefore, the research question is whether this differentiation can be observed in the intercepted encrypted traffic.

\begin{figure}
    \begin{small}
    \begin{verbatim}
; <<>> DiG 9.11.3-1ubuntu1.2-Ubuntu <<>> www.google.com
;; global options: +cmd
;; Got answer:
;; ->>HEADER<<- opcode: QUERY, status: NOERROR, id: 18417
;; flags: qr rd ra; QUERY: 1, ANSWER: 1, AUTHORITY: 0, ADDITIONAL: 1

;; OPT PSEUDOSECTION:
; EDNS: version: 0, flags:; udp: 65494
;; QUESTION SECTION:
;www.google.com.			IN	A

;; ANSWER SECTION:
www.google.com.		238	IN	A	216.58.205.164

;; Query time: 44 msec
;; SERVER: 127.0.0.53#53(127.0.0.53)
;; WHEN: Thu Oct 25 01:08:12 EEST 2018
;; MSG SIZE  rcvd: 59
    \end{verbatim}
    \end{small}
    \caption{The DNS response when querying for Google.com.}
    \label{fig:dig_Google}
\end{figure}

\begin{figure}
    \begin{small}
    \begin{verbatim}
; <<>> DiG 9.11.3-1ubuntu1.2-Ubuntu <<>> rejuxlip.ru
;; global options: +cmd
;; Got answer:
;; ->>HEADER<<- opcode: QUERY, status: NXDOMAIN, id: 64200
;; flags: qr rd ra; QUERY: 1, ANSWER: 0, AUTHORITY: 0, ADDITIONAL: 1

;; OPT PSEUDOSECTION:
; EDNS: version: 0, flags:; udp: 65494
;; QUESTION SECTION:
;rejuxlip.ru.			IN	A

;; Query time: 2475 msec
;; SERVER: 127.0.0.53#53(127.0.0.53)
;; WHEN: Thu Oct 25 01:10:58 EEST 2018
;; MSG SIZE  rcvd: 40
    \end{verbatim}
    \end{small}
    \caption{The DNS response when querying for rejuxlip.ru.}
    \label{fig:dig_DGA}
\end{figure}

\section{Discussion}
\label{sec:discussion}
In the following paragraphs we discuss our findings of our experimental results. To this end, we first discuss the experimental results using traffic analysis and then we present how one can efficiently perform bot attribution.

\subsection{Traffic analysis}
\begin{figure}[th!]
    \centering
    \begin{subfigure}{0.32\textwidth}
        \centering
        \includegraphics[width=\linewidth]{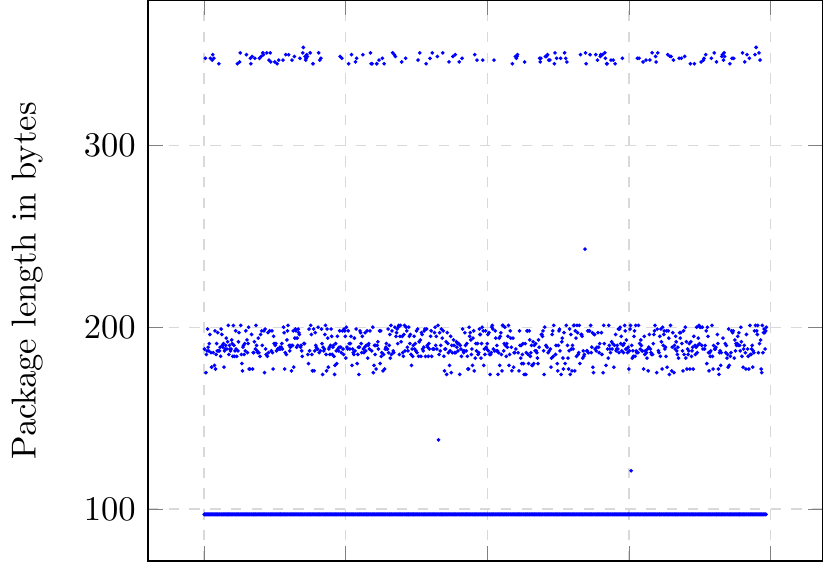}
        \caption{Conficker}
        \label{fig:conflicker_tls}
    \end{subfigure}
    \begin{subfigure}{0.32\textwidth}
        \centering
        \includegraphics[width=\linewidth]{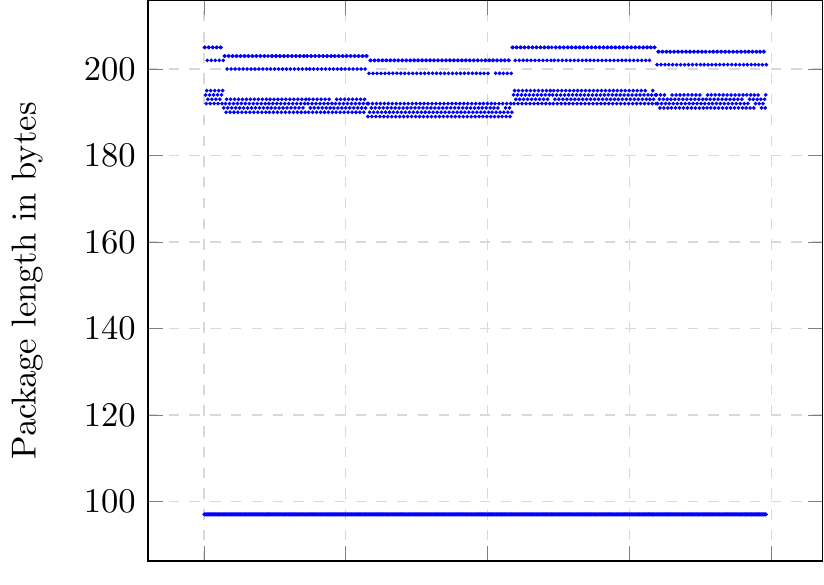}
        \caption{CryptoLocker}
        \label{fig:CryptoLocker_tls}
    \end{subfigure}
    \begin{subfigure}{0.32\textwidth}
        \centering
        \includegraphics[width=\linewidth]{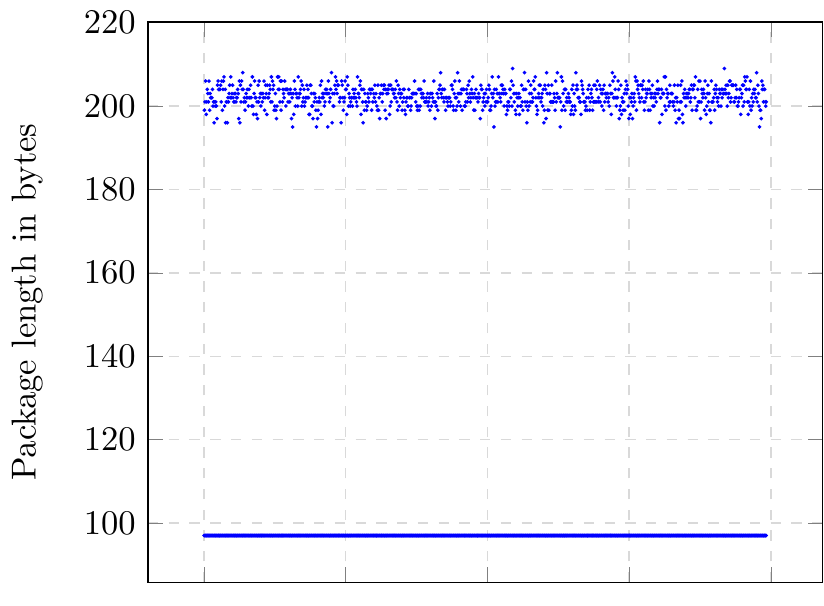}
        \caption{GOZ}
        \label{fig:GOZ_tls}
    \end{subfigure}
    
    \begin{subfigure}{0.32\textwidth}
        \centering
        \includegraphics[width=\linewidth]{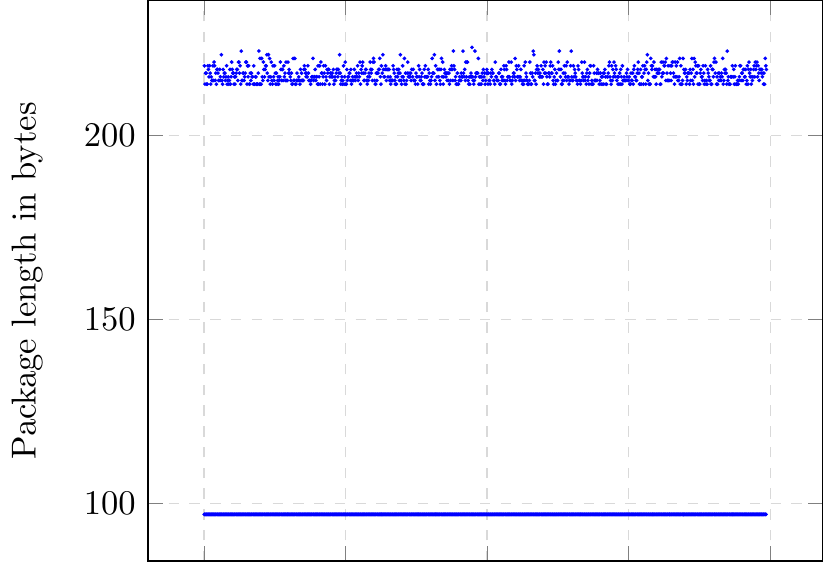}
        \caption{Matsnu}
        \label{fig:Matsnu_tls}
    \end{subfigure}
    \begin{subfigure}{0.32\textwidth}
        \centering
        \includegraphics[width=\linewidth]{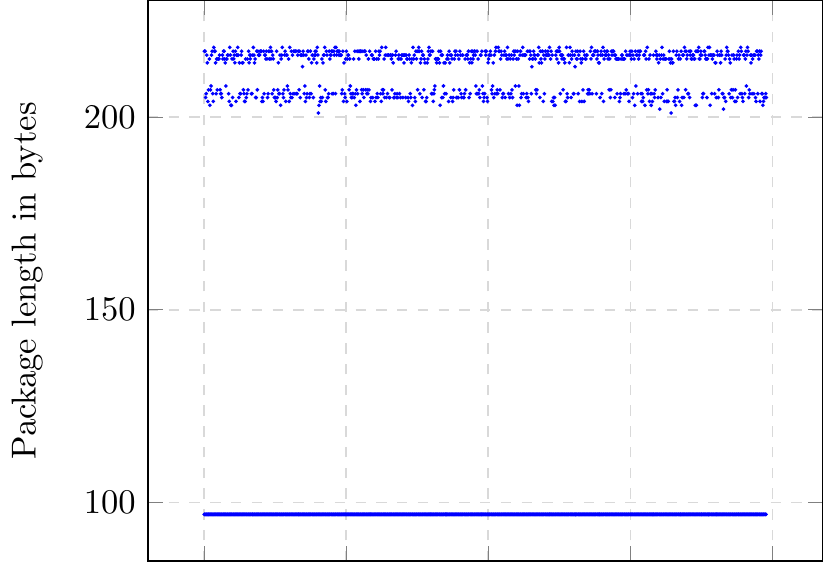}
        \caption{New GOZ.}
        \label{fig:newGOZ_tls}
    \end{subfigure}
    \begin{subfigure}{0.32\textwidth}
        \centering
        \includegraphics[width=\linewidth]{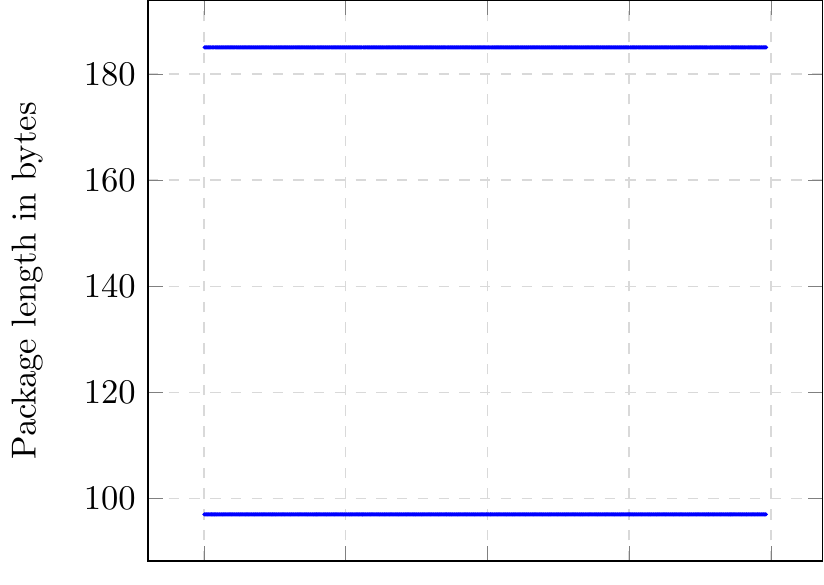}
        \caption{Pushdo}
        \label{fig:Pushdo_tls}
    \end{subfigure}
    
    \begin{subfigure}{0.32\textwidth}
        \centering
        \includegraphics[width=\linewidth]{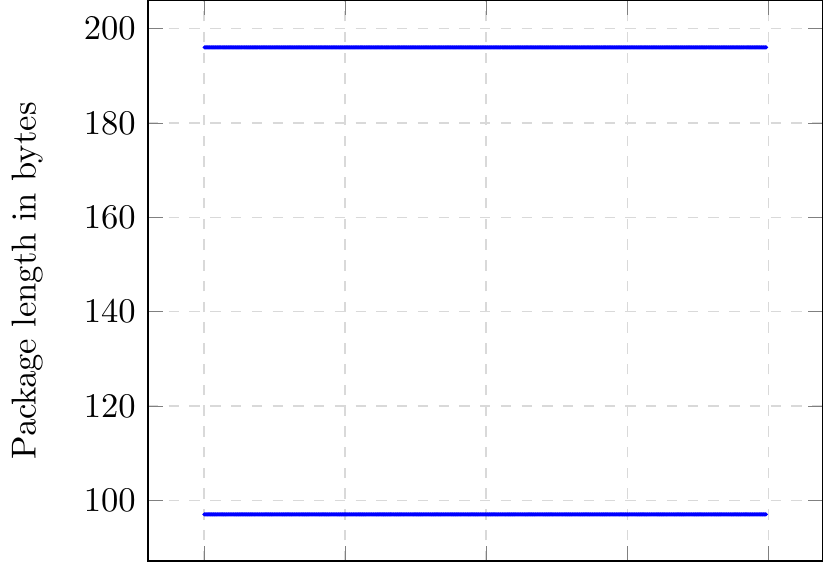}
        \caption{Ramdo}
        \label{fig:Ramdo_tls}
    \end{subfigure}
    \begin{subfigure}{0.32\textwidth}
        \centering
        \includegraphics[width=\linewidth]{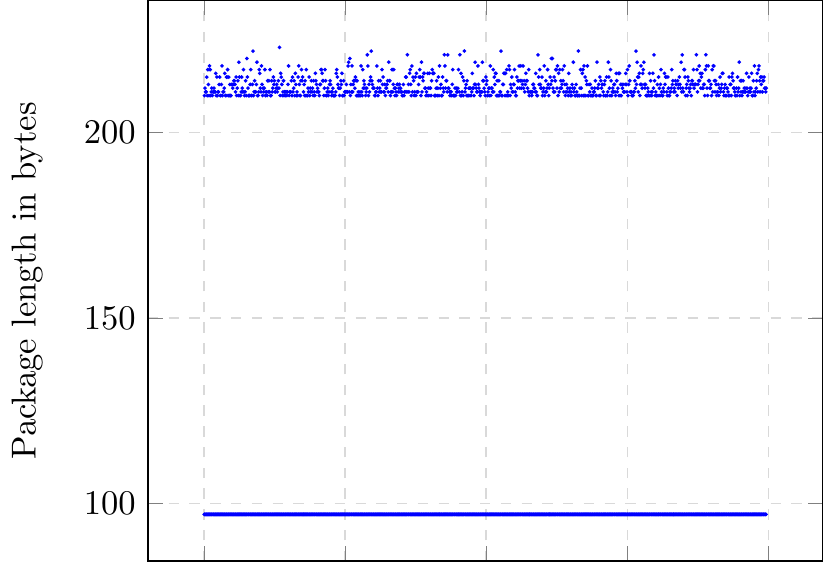}
        \caption{Rovnix}
        \label{fig:Rovnix_tls}
    \end{subfigure}
    \begin{subfigure}{0.32\textwidth}
        \centering
        \includegraphics[width=\linewidth]{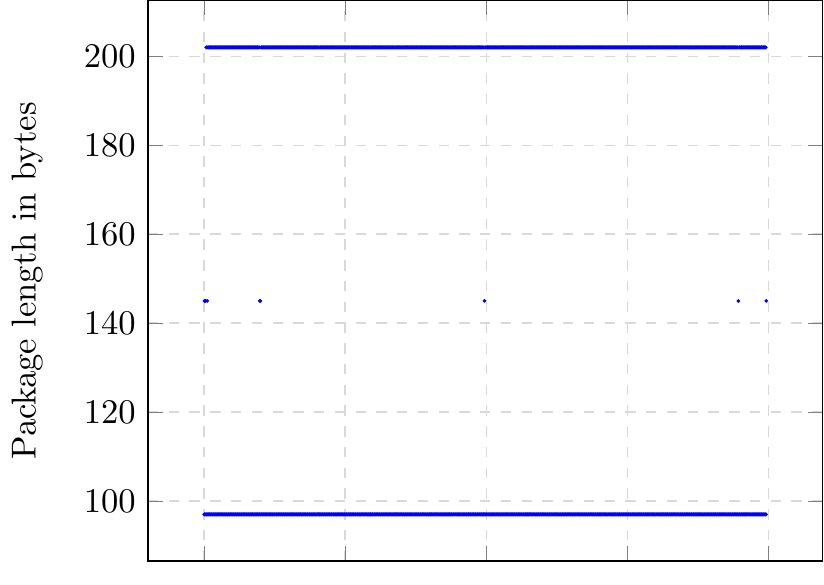}
        \caption{Tinba}
        \label{fig:Tinba_tls}
    \end{subfigure}
    
    \begin{subfigure}{0.32\textwidth}
        \centering
        \includegraphics[width=\linewidth]{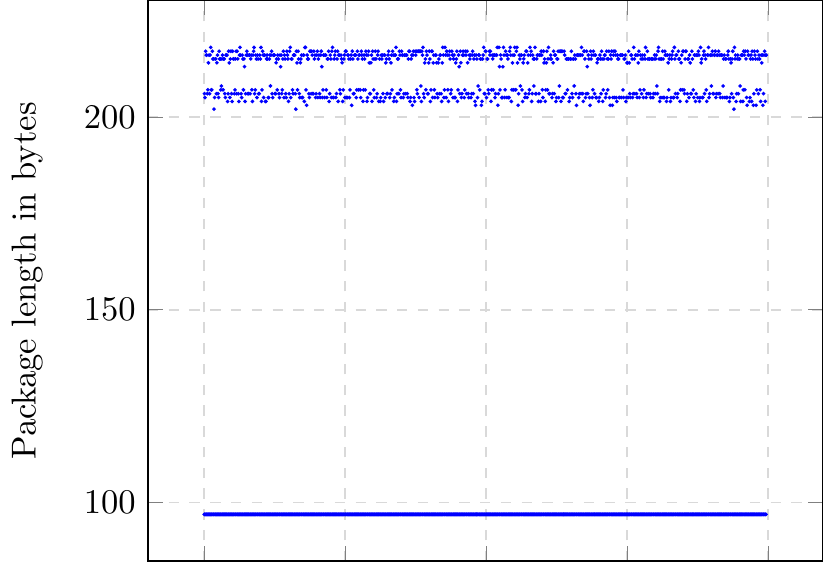}
        \caption{Zeus}
        \label{fig:Zeus_tls}
    \end{subfigure}
    \begin{subfigure}{0.32\textwidth}
        \centering
        \includegraphics[width=\linewidth]{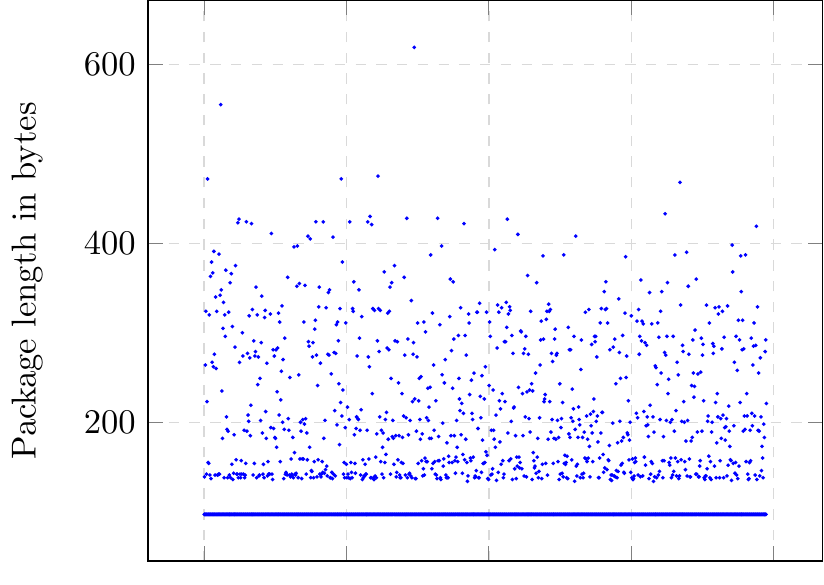}
        \caption{Alexa}
        \label{fig:alexa_tls}
    \end{subfigure}
    \caption{Package size diversity for different datasets (DGAs and Alexa) using DNS over TLS.}
    \label{fig:packet_visualisation_tls}
\end{figure}

\begin{figure}[th!]
    \centering
    \begin{subfigure}{0.32\textwidth}
        \centering
        \includegraphics[width=\linewidth]{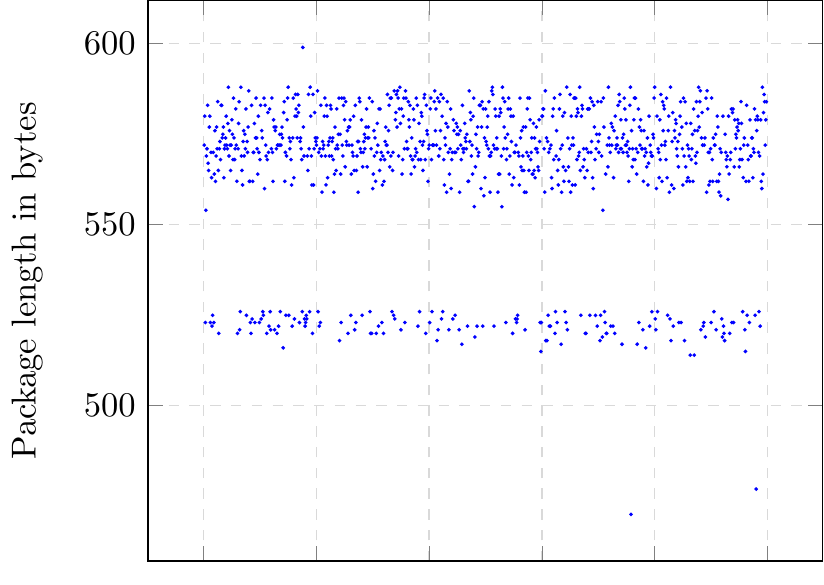}
        \caption{Conflicker}
        \label{fig:conflicker_https}
    \end{subfigure}
    \begin{subfigure}{0.32\textwidth}
        \centering
        \includegraphics[width=\linewidth]{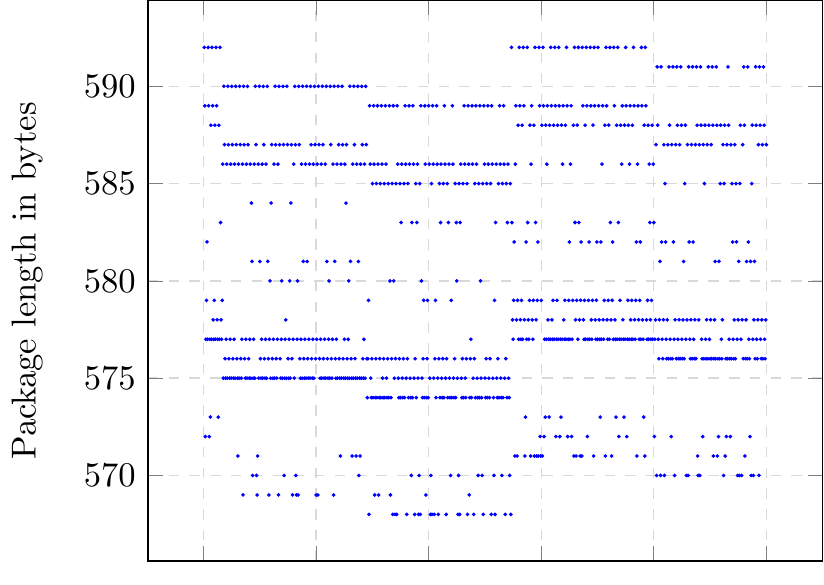}
        \caption{CryptoLocker}
        \label{fig:CryptoLocker_https}
    \end{subfigure}
    \begin{subfigure}{0.32\textwidth}
        \centering
        \includegraphics[width=\linewidth]{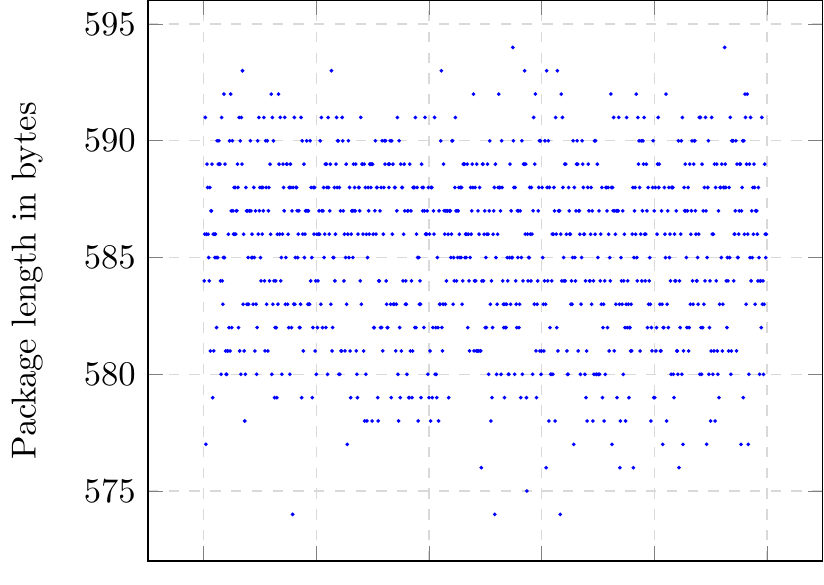}
        \caption{GOZ}
        \label{fig:GOZ_https}
    \end{subfigure}
    
    \begin{subfigure}{0.32\textwidth}
        \centering
        \includegraphics[width=\linewidth]{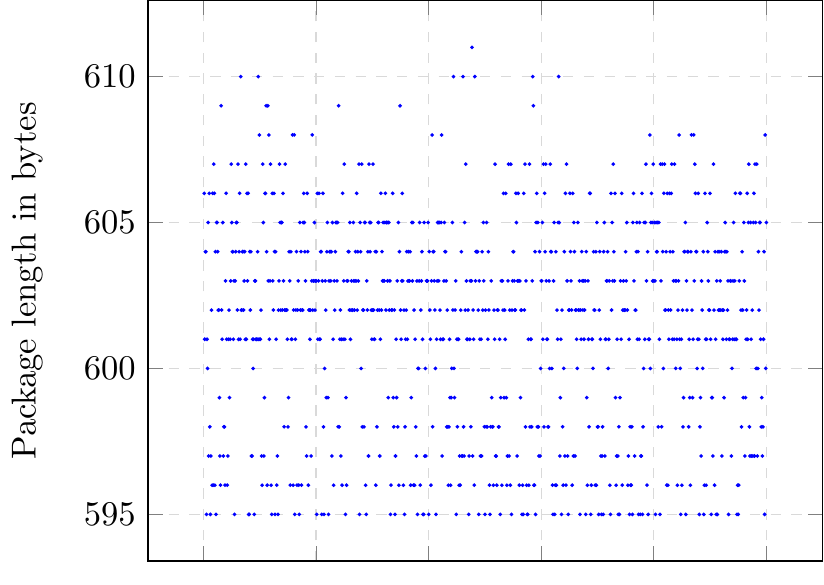}
        \caption{Matsnu}
        \label{fig:Matsnu_https}
    \end{subfigure}
    \begin{subfigure}{0.32\textwidth}
        \centering
        \includegraphics[width=\linewidth]{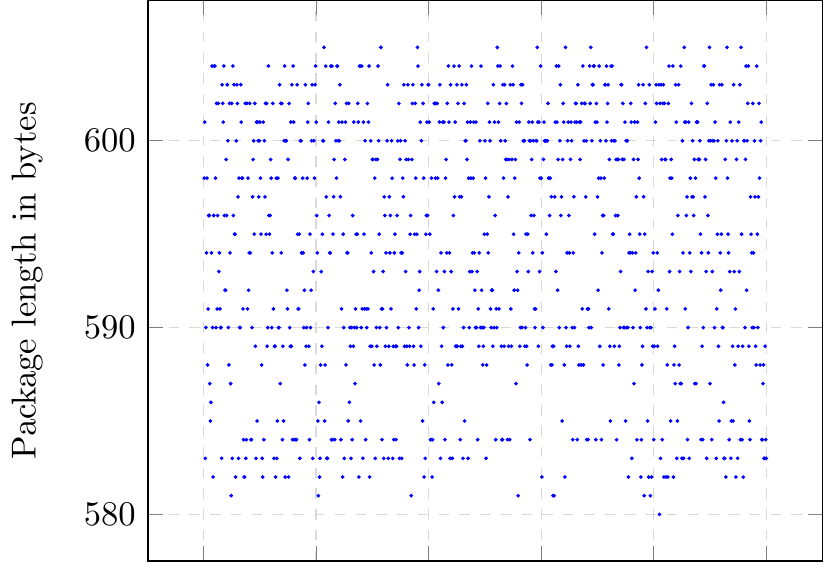}
        \caption{new GOZ}
        \label{fig:newGOZ_https}
    \end{subfigure}
    \begin{subfigure}{0.32\textwidth}
        \centering
        \includegraphics[width=\linewidth]{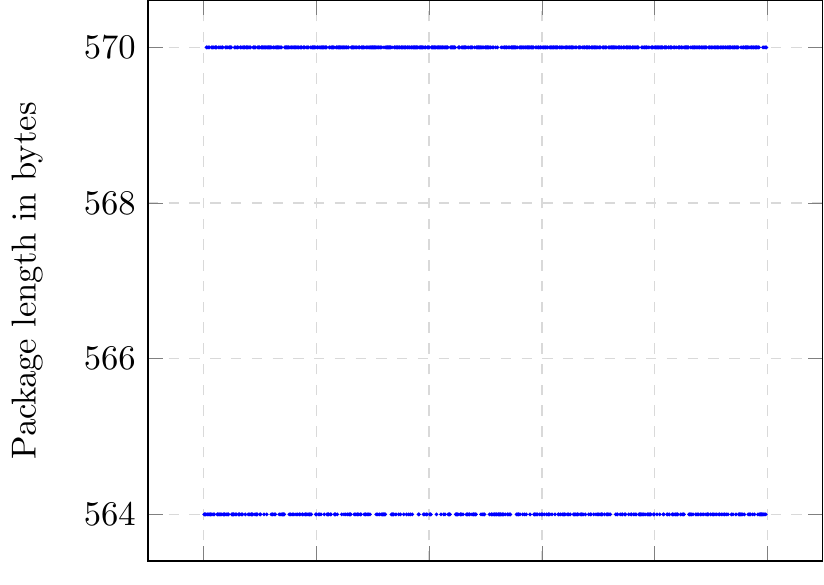}
        \caption{Pushdo}
        \label{fig:Pushdo_https}
    \end{subfigure}
    
    \begin{subfigure}{0.32\textwidth}
        \centering
        \includegraphics[width=\linewidth]{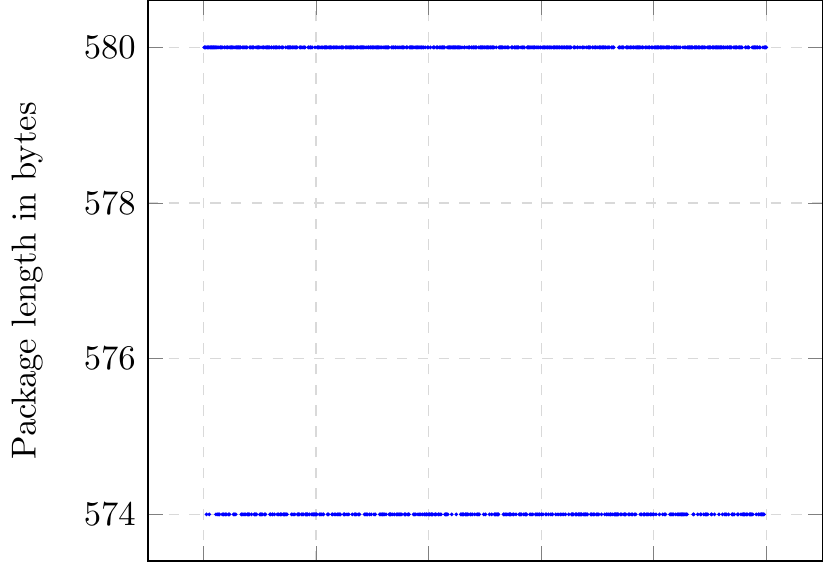}
        \caption{Ramdo}
        \label{fig:Ramdo_https}
    \end{subfigure}
    \begin{subfigure}{0.32\textwidth}
        \centering
        \includegraphics[width=\linewidth]{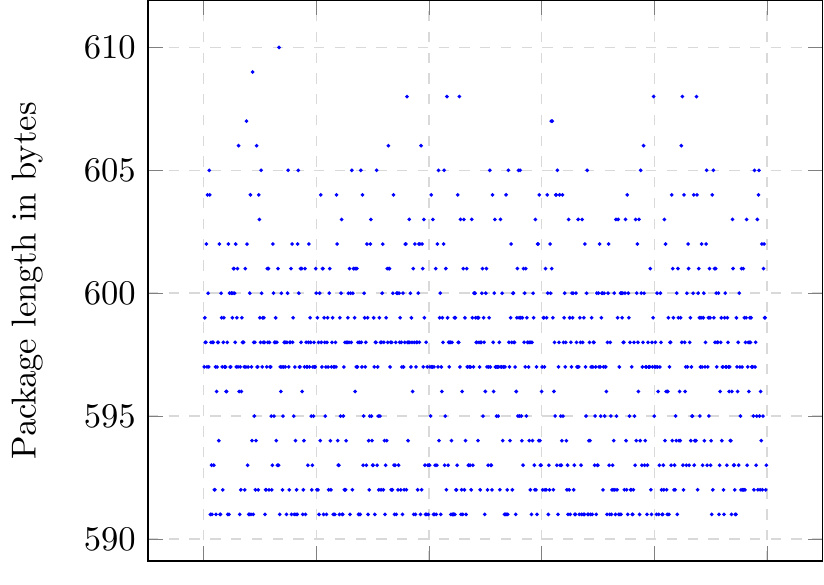}
        \caption{Rovnix}
        \label{fig:Rovnix}
    \end{subfigure}
    \begin{subfigure}{0.32\textwidth}
        \centering
        \includegraphics[width=\linewidth]{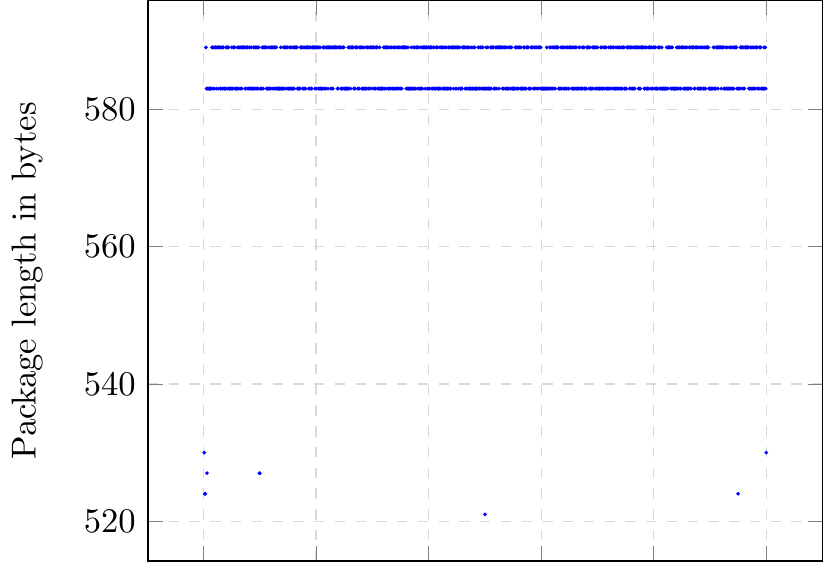}
        \caption{Tinba}
        \label{fig:Tinba_https}
    \end{subfigure}
    
    \begin{subfigure}{0.32\textwidth}
        \centering
        \includegraphics[width=\linewidth]{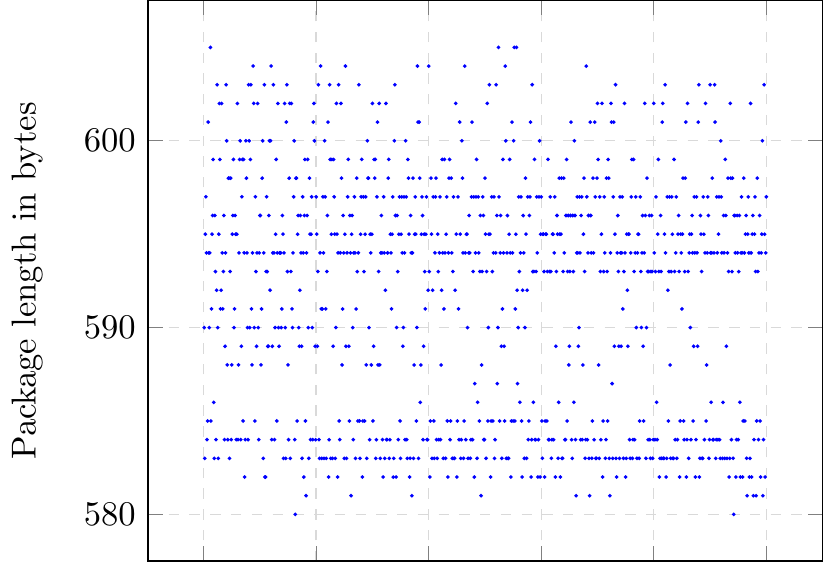}
        \caption{Zeus}
        \label{fig:Zeus_https}
    \end{subfigure}
    \begin{subfigure}{0.32\textwidth}
        \centering
        \includegraphics[width=\linewidth]{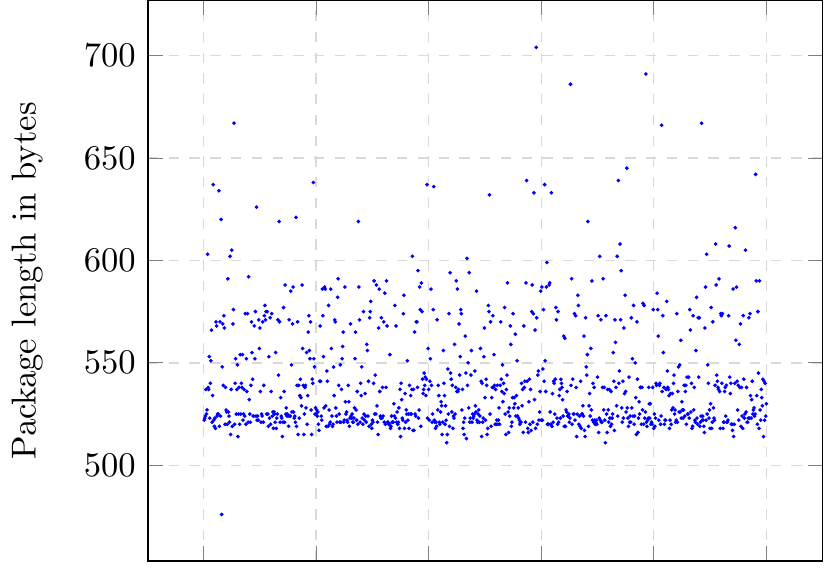}
        \caption{Alexa}
        \label{fig:alexa_https}
    \end{subfigure}
    \caption{Package size diversity for different datasets (DGAs and Alexa) using DNS over https.}
    \label{fig:packet_visualisation_https}
\end{figure}

On a macroscopic level, examining the size of the responses in each case we notice that some interesting patterns emerge, see Figures \ref{fig:conflicker_tls}-\ref{fig:alexa_tls} and \ref{fig:conflicker_https}-\ref{fig:alexa_https}. It should be noted that due to its implementation when using DNS over TLS, each response consists of two packets with the first one having a length of 97 bytes. Evidently, DGAs which produce domains with static length, end up having server responses with static length as well, see Figures \ref{fig:Pushdo_tls}, \ref{fig:Pushdo_https}, \ref{fig:Ramdo_tls},\ref{fig:Ramdo_https}, \ref{fig:Tinba_tls} and \ref{fig:Tinba_https}. Similarly, DGAs which produce domains with little variance on their length, see \ref{fig:CryptoLocker_tls} and \ref{fig:CryptoLocker_https}, have responses with little variance in their length. Clearly, the Alexa dataset presents the most significant variance in the response lengths as it has the most extensive diversity in terms of both lengths of domains and the response is expected to contain several values for the IPs of the hosts. In Table \ref{tbl:stats} we provide an overview of the statistical properties of the responses in each case. More precisely, columns Min, Max, Average and Stdev refer to the corresponding measures for the packet size of each dataset. Unique values indicate whether all possible sizes of packets are represented in each dataset. For instance, in Pushdo for DNS over HTTPS, while the range is from 564 to 570, only two values are found in our experiments.


It is clear that the results mentioned above indicate that even though the DNS queries are performed covertly via legitimate DNS servers, traffic analysis can efficiently determine the queries on existing and NXDomains with almost 100\% accuracy if the DGA generates domains of static length. In this regard, one could develop a lightweight approach to constructing IoCs. The corresponding rule would look for packages to privacy-preserving DNS servers like Google, Cloudflare etc. over HTTPS and then look for the 97 vs X pattern or the stable value pattern on the received packages to determine whether a device has been compromised and tries to perform DNS queries over a covert channel.

\begin{table}[th!]
\centering
\begin{tabular}{llrrrrr}\toprule
 & \textbf{Dataset} & \textbf{Min} & \textbf{Max} & \textbf{Average} & \textbf{Stdev} & \textbf{Unique values} \\\midrule
\multirow{11}{*}{\textbf{HTTPS}} & alexa & 476 & 704 & 541.610 & 27.646 & 109\\
& Conflicker & 470 & 599 & 565.162 & 20.612 & 50\\
& CryptoLocker & 568 & 592 & 580.045 & 6.586 & 25\\
& GOZ & 574 & 594 & 585.218 & 3.710 & 21\\
& Matsnu & 595 & 611 & 601.425 & 3.617 & 17\\
& new GOZ & 580 & 605 & 594.347 & 6.797 & 26\\
& Pushdo & 564 & 570 & 567.806 & 2.890 & 2\\
& Ramdo & 574 & 580 & 577.708 & 2.915 & 2\\
& Rovnix & 591 & 610 & 596.900 & 3.979 & 20\\
& Tinba & 521 & 589 & 585.445 & 6.410 & 6\\
& Zeus & 580 & 605 & 591.737 & 6.414 & 26\\\midrule
\multirow{11}{*}{\textbf{TLS}} & alexa & 134 & 619 & 220.951 & 80.746 & 235\\
& Conflicker & 121 & 354 & 214.670 & 58.580 & 37\\
& CryptoLocker & 189 & 205 & 196.532 & 5.484 & 14\\
& GOZ & 195 & 209 & 202.217 & 2.484 & 15\\
& Matsnu & 214 & 224 & 216.566 & 2.046 & 11\\
& new GOZ & 201 & 218 & 211.854 & 5.281 & 14\\
& Pushdo & 185 & 185 & 185.000 & 0.000 & 1\\
& Ramdo & 196 & 196 & 196.000 & 0.000 & 1\\
& Rovnix & 210 & 223 & 212.759 & 2.684 & 14\\
& Tinba & 145 & 202 & 201.483 & 5.405 & 2\\
& Zeus & 202 & 218 & 211.520 & 5.245 & 13\\\bottomrule
\end{tabular}
\caption{Statistical properties. Note that for TLS we omit the packages with length 97 bytes as this appears in all interactions.}
\label{tbl:stats}
\end{table}


However, a thorough and complete detection requires providing attribution information which in this case it would be threat intelligence on the type of infection and bot attempting to communicate with its C\&C or botmaster. Detecting NXDomain responses is expected to be performed at an early stage of the intrusion detection process to allow the security controls to block the connection or respond in the prescribed security policy manner. The incident response process would be considered complete if it would be capable of providing evidence of the bot that has successfully compromised the system. As such, the aim of the incident response exercise is twofold: a) identify quickly that a bot has infected the system; b) identify the particular bot or family. 

Currently, in an unencrypted, standard DNS setting, the computed IoC describing a bot infection would be a large number of NXDomain responses within a particular time frame. We have described above how to replicate this IoC under TLS and HTTPS communication. In the following section, we present an approach for constructing higher quality IoCs capable of identifying the underlying bot.

\subsection{Bot attribution}
The approach for constructing bot IoCs is based on the following assumptions:
\begin{itemize}
    \item the bot uses a fixed DGA strategy to generate the domains,
    \item the domains are generated sequentially,
    \item the process is repeated until an existing domain is received.
\end{itemize}

These assumptions allow us to consider the network traffic generated by the bot as a time series variable. This is also the first significant distinction between the Alexa dataset and all other bots we considered in our experiments. In what follows, we present only the results for the case of DNS over HTTPS traffic, yet similar results apply for DNS over TLS. 

\begin{figure}[th!]
    \centering
    \begin{subfigure}{0.32\textwidth}
        \centering
        \includegraphics[width=\linewidth]{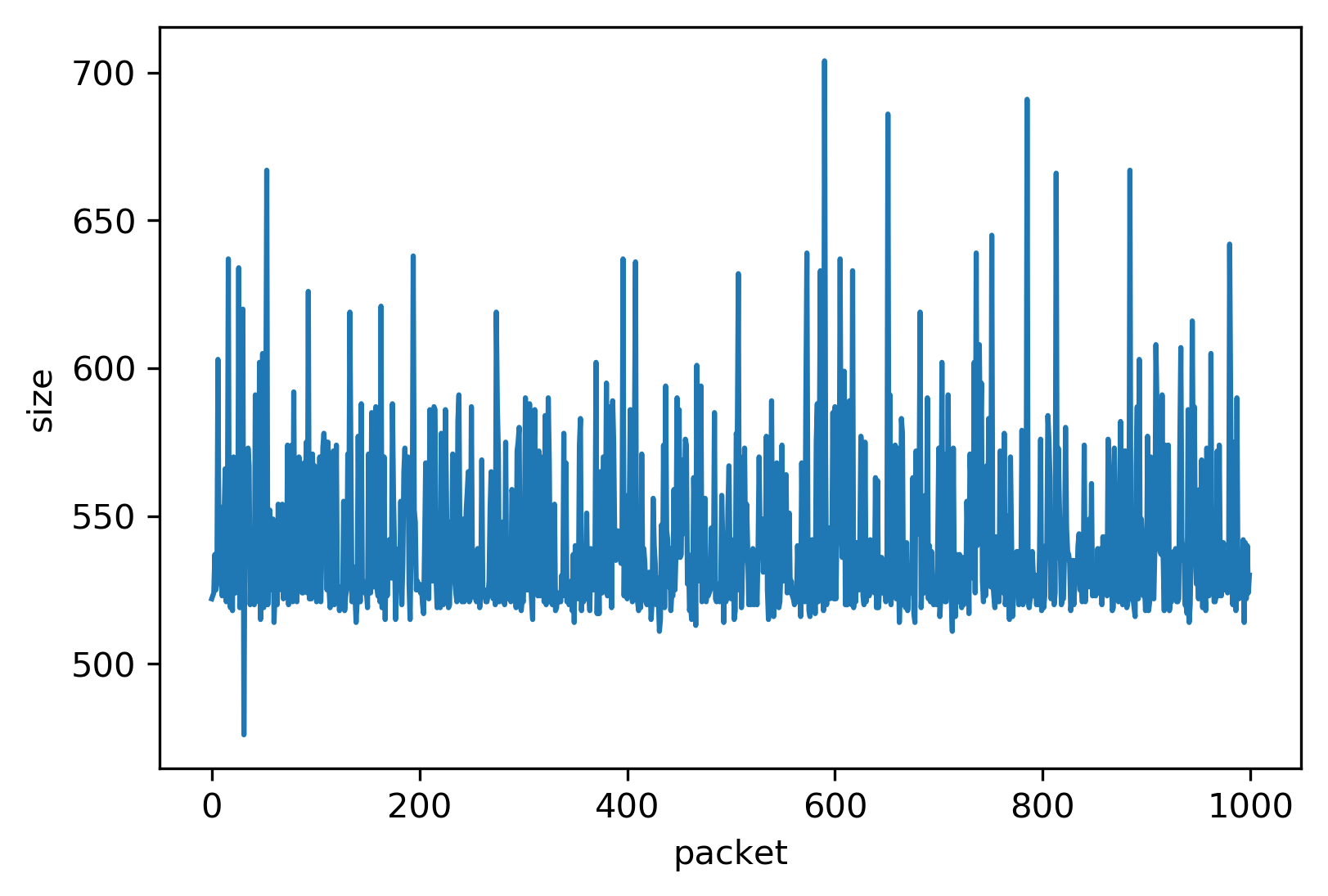}
        \caption{Alexa}
    \end{subfigure}~
        \begin{subfigure}{0.32\textwidth}
        \centering
        \includegraphics[width=\linewidth]{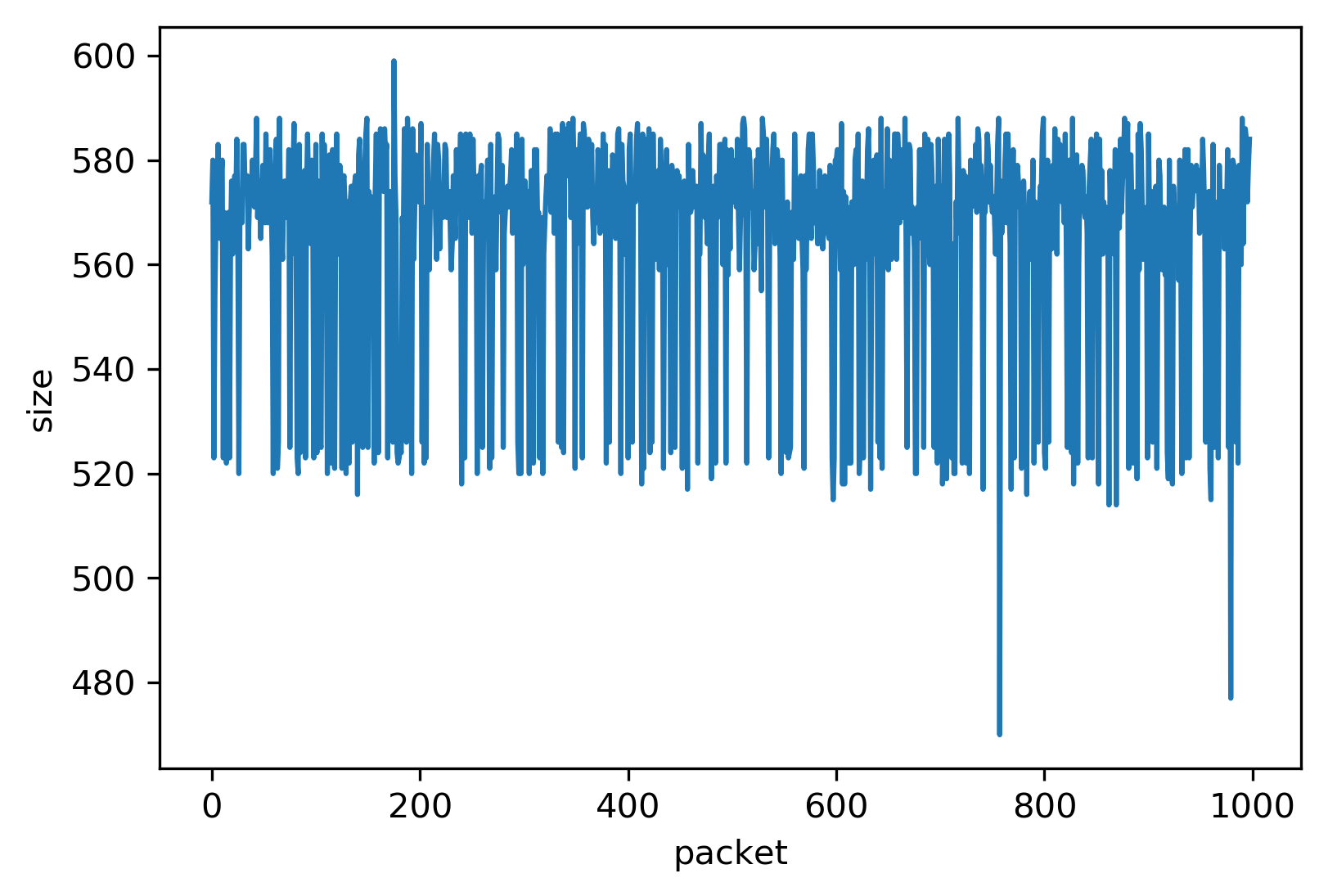}
        \caption{Conficker}
    \end{subfigure}~
    \begin{subfigure}{0.32\textwidth}
        \centering
        \includegraphics[width=\linewidth]{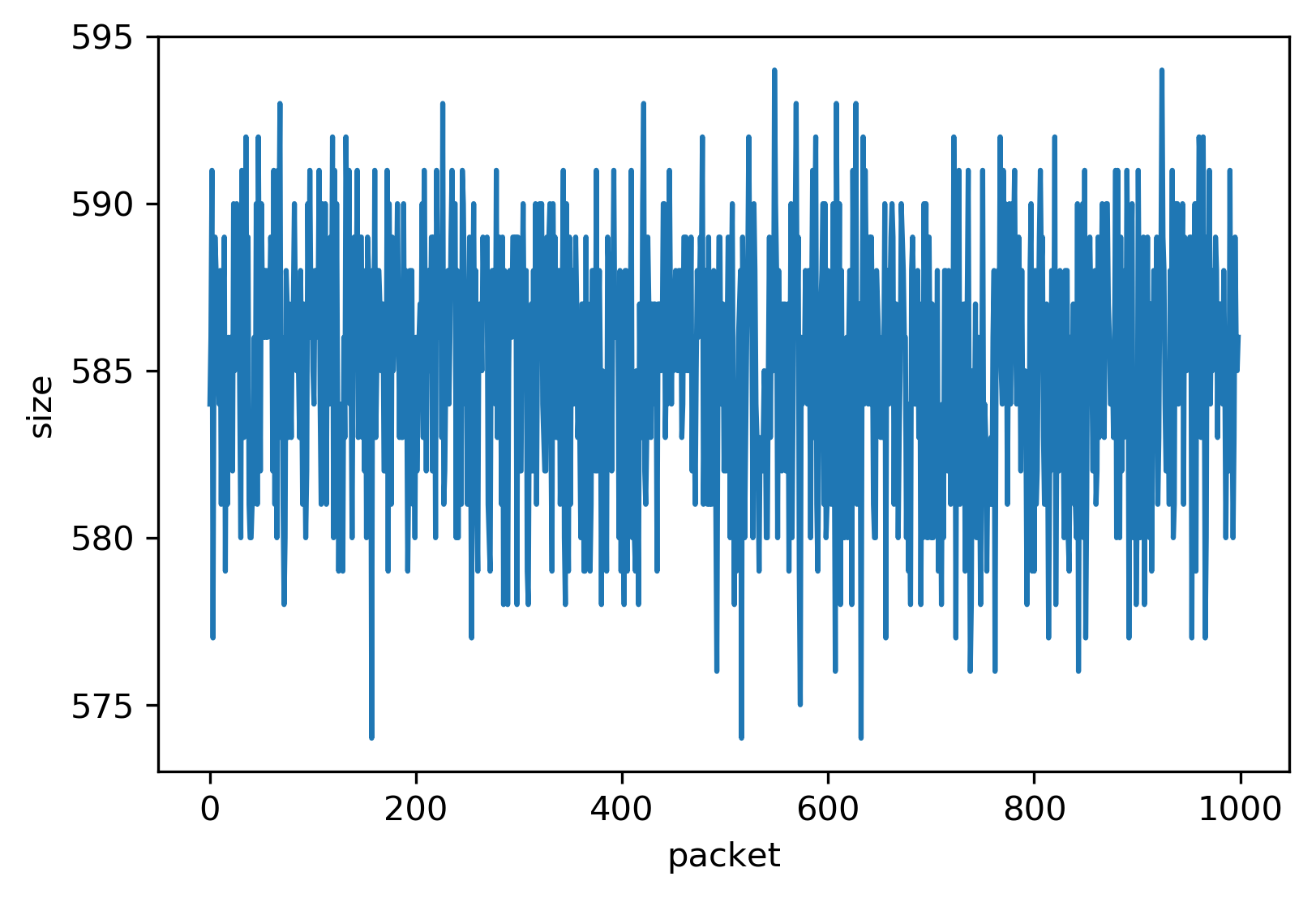}
        \caption{Goz}
    \end{subfigure}
\caption{Time plots of Alexa, Conficker and Goz}
\label{fig:time}
\end{figure}

Figure \ref{fig:time} shows a sample of the datasets, more precisely Alexa, Conficker and Goz, plotted over time. More precisely, the figure illustrates the packet size (axis Y) for each dataset, if we consider that one performed a covert DNS query for each of the domains of the corresponding dataset sequentially (axis X). It can be observed that the variables display a significant amount of noise.

In the proposed approach we attempt to apply filtering in order to remove the noise and identify any trend. This approach is commonplace in the economic analysis of time series, and we explore its applicability to the current problem domain. More specifically, in the context of time series analysis the objective is to identify a trend so that this can be used to make predictions of the future values of a given variable. This can be attempted if one would establish that there is correlation between the immediate past observations (lags) for any given point in the series. If this is the case, this knowledge can be extracted from the series in a form of a trend, thus breaking down the series (or \emph{signal}) to a linear combination of a \emph{trend} and a \emph{cycle}, with the latter capturing any periodicity. Once these trends are identified, we attempt to see whether they are distinct for each malware and different from benign traffic. Moreover the examined datasets can be considered to be time series (although the $x$ axis does not explicitly capture time) as the DNS packets generated from a DGA follow a sequential order over time.

\begin{figure}[th!]
    \centering
    \begin{subfigure}{0.32\textwidth}
        \centering
        \includegraphics[width=\linewidth]{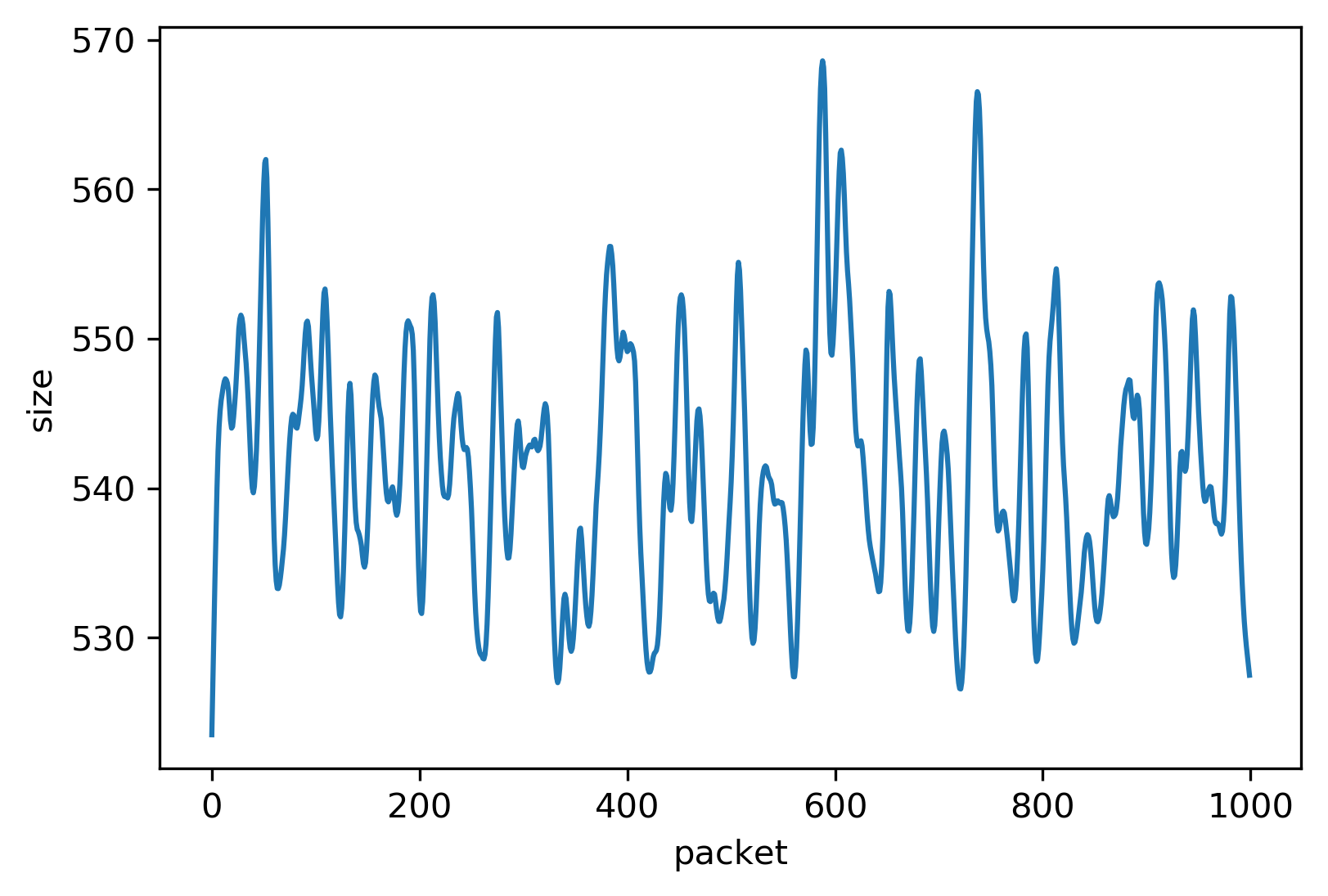}
        \caption{Alexa's trend}
        \label{fig:alexa_trend}
    \end{subfigure}
        \begin{subfigure}{0.32\textwidth}
        \centering
        \includegraphics[width=\linewidth]{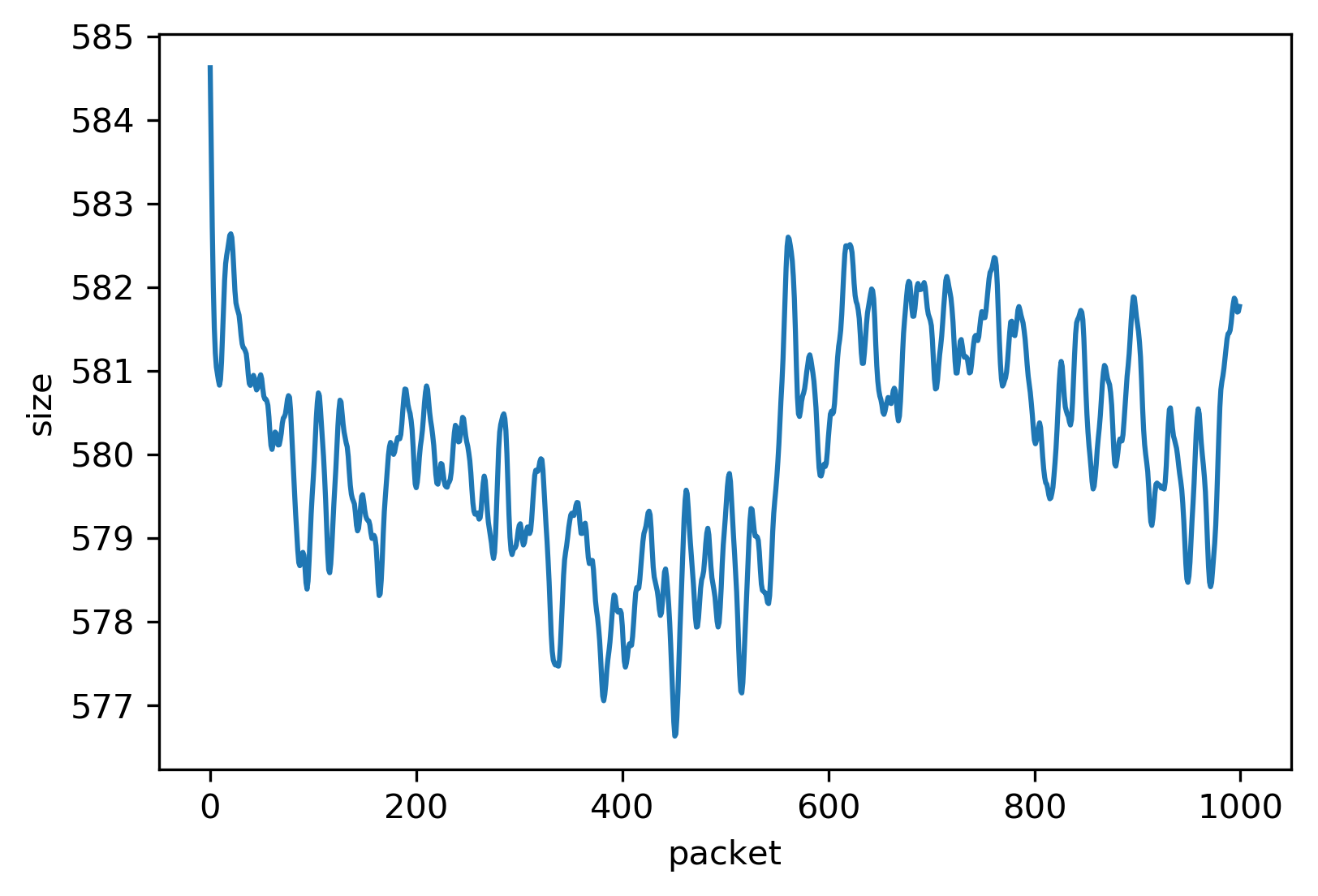}
        \caption{Cryptolocker's trend}
        \label{fig:crypt_trend}
    \end{subfigure}
    \begin{subfigure}{0.32\textwidth}
        \centering
        \includegraphics[width=\linewidth]{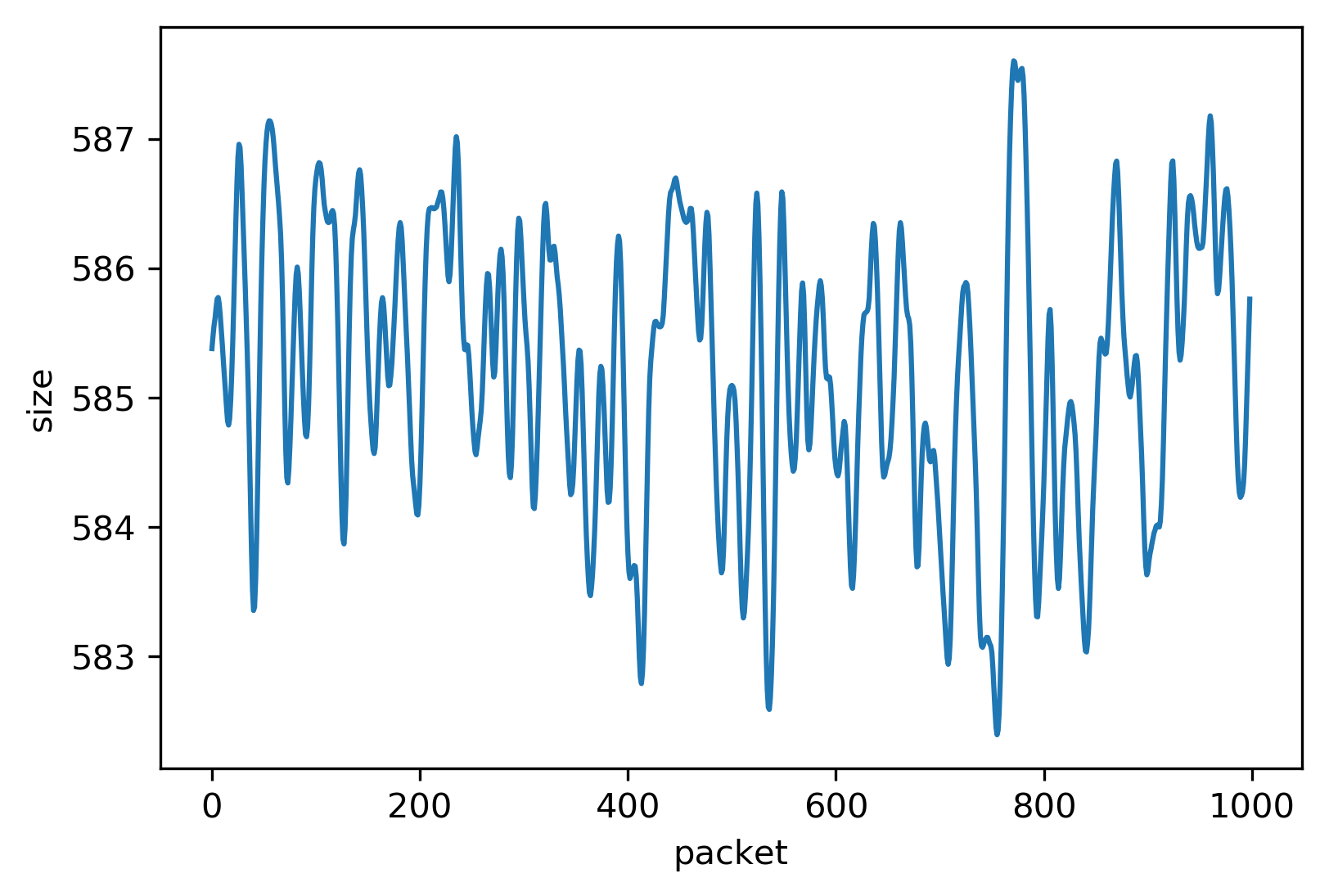}
        \caption{Goz's trend}
        \label{fig:goz_trend}
    \end{subfigure}
        \begin{subfigure}{0.32\textwidth}
        \centering
        \includegraphics[width=\linewidth]{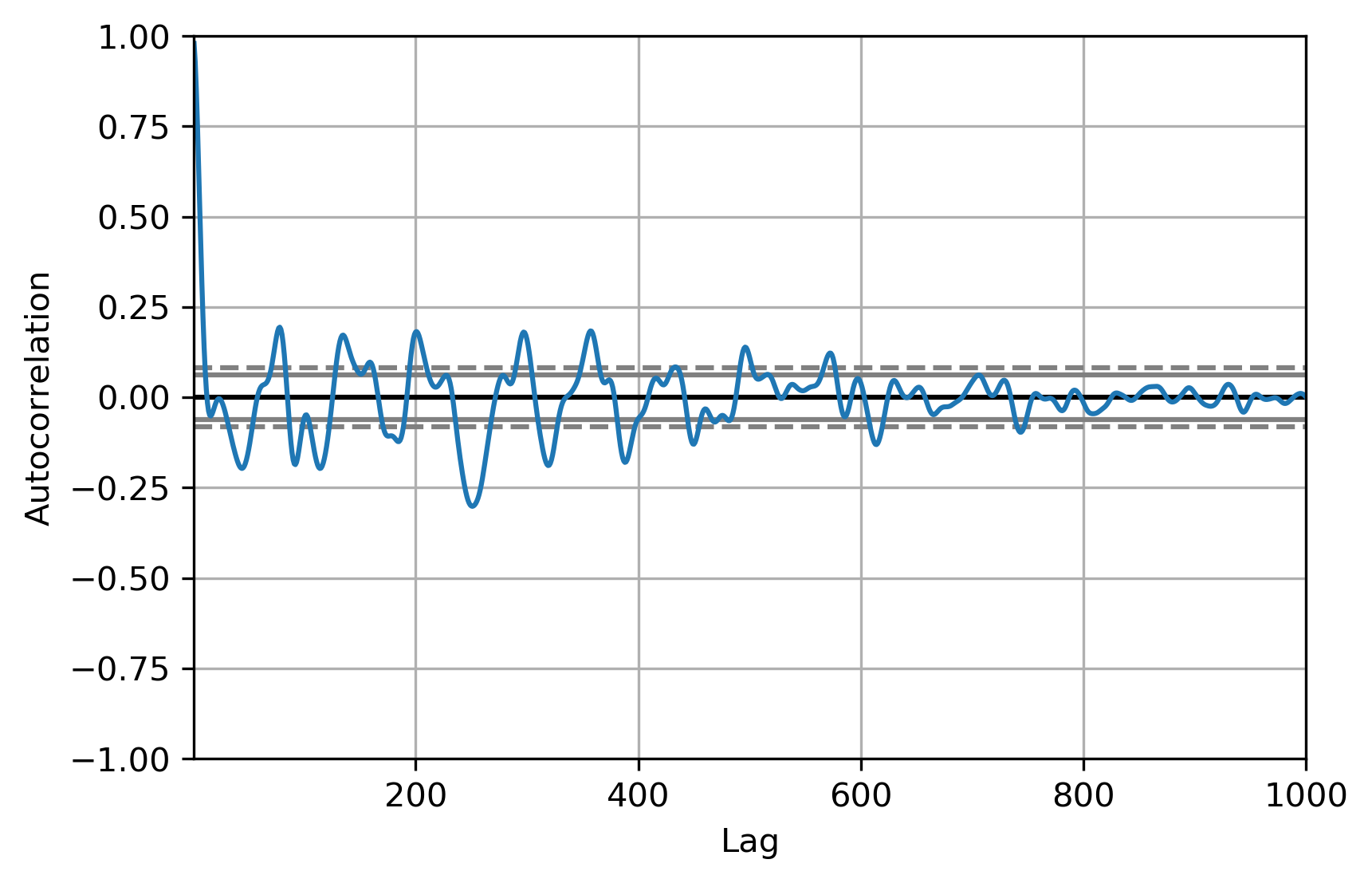}
        \caption{Alexa's trend autocorrelation}
        \label{fig:alexa_trend_auto}
    \end{subfigure}
        \begin{subfigure}{0.32\textwidth}
        \centering
        \includegraphics[width=\linewidth]{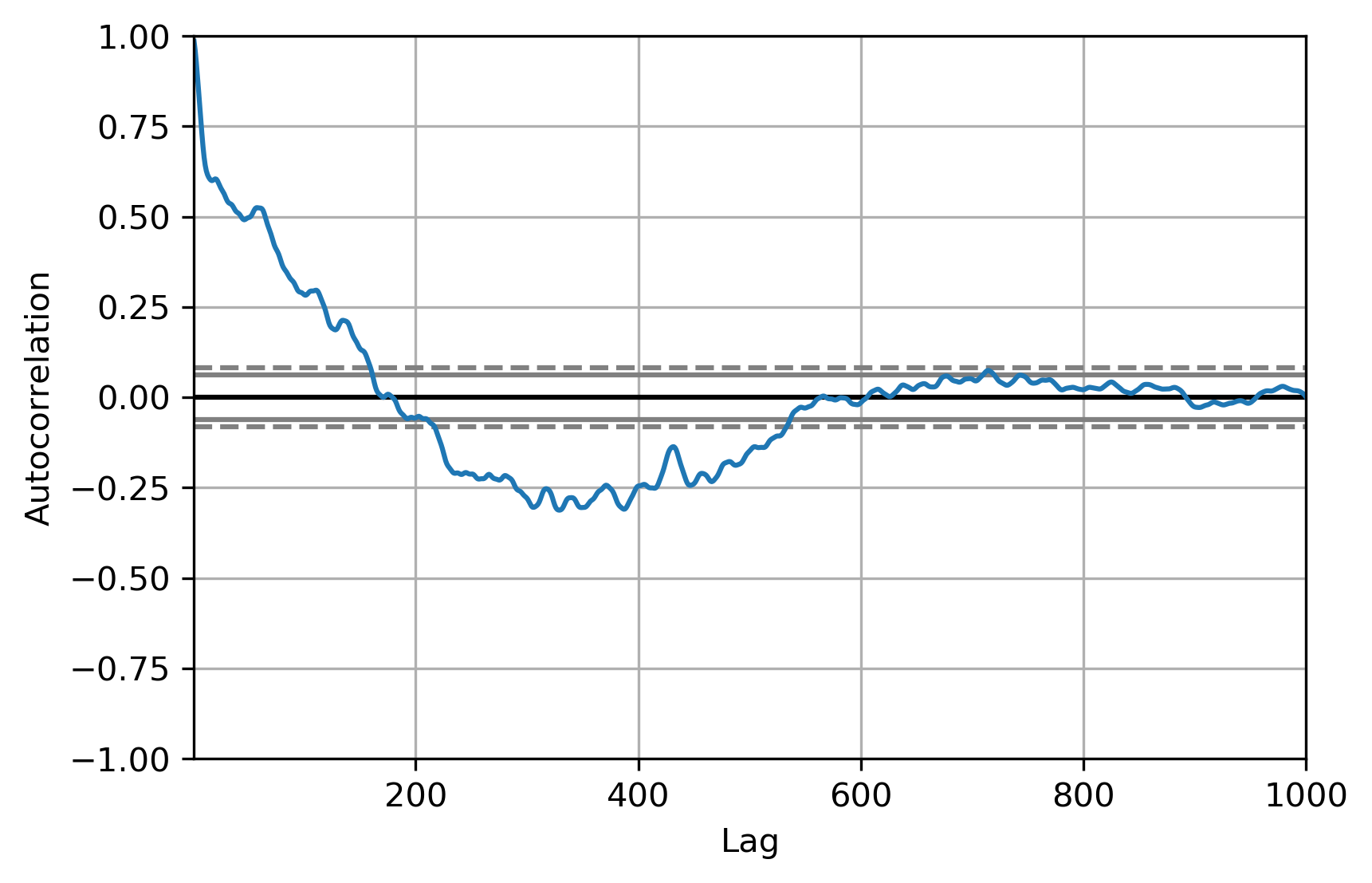}
        \caption{Cryptolocker's trend autocorrelation}
        \label{fig:crypt_trend_auto}
    \end{subfigure}
    \begin{subfigure}{0.32\textwidth}
        \centering
        \includegraphics[width=\linewidth]{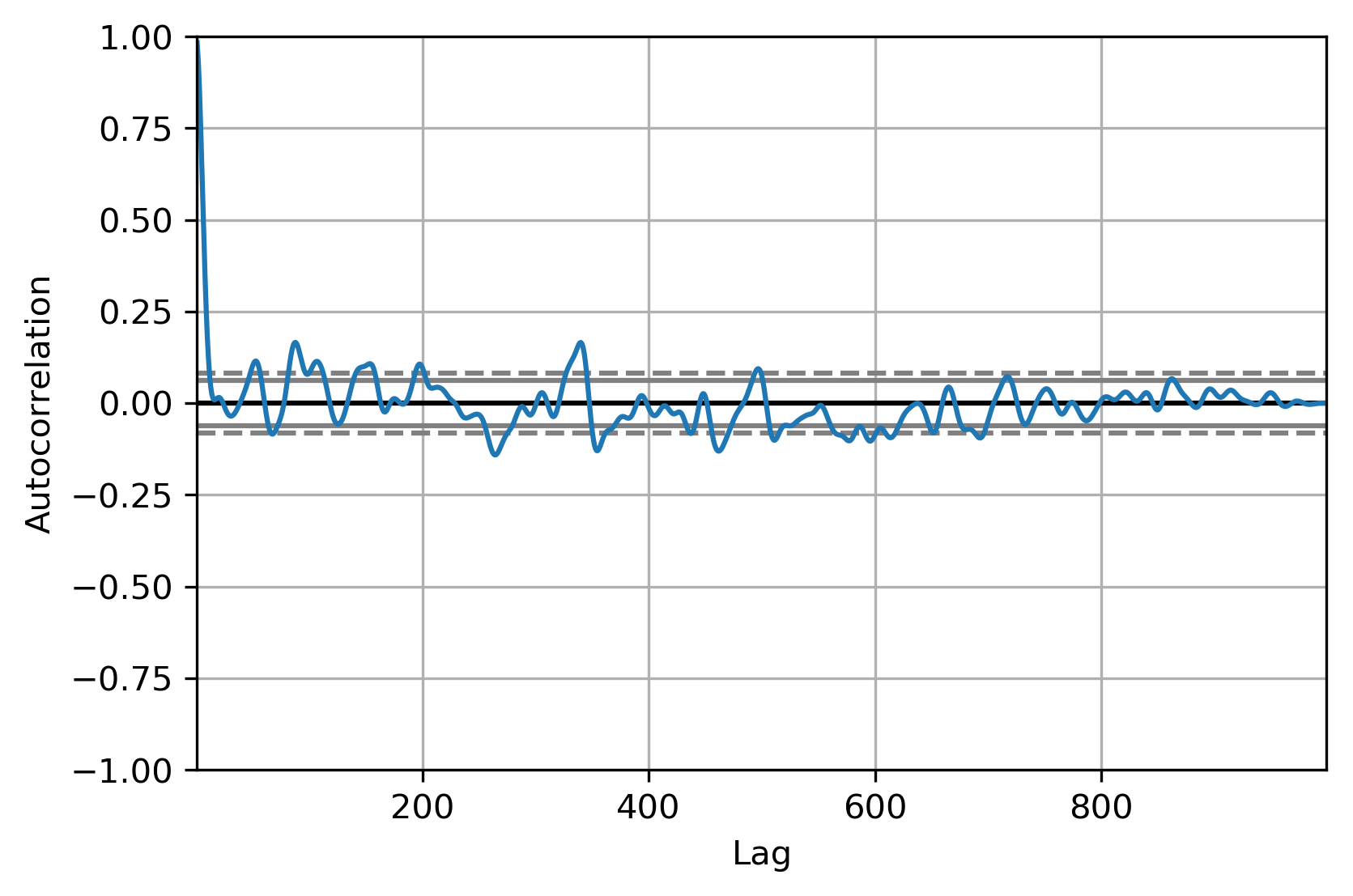}
        \caption{Goz's trend autocorrelation}
        \label{fig:goz_trend_auto}
    \end{subfigure}
\caption{Trends and autocorrelations}
\label{fig:trend}
\end{figure}

The approach of time series analysis is as follows. Assuming that the variable is composed of a cycle (which is true as we have DNS requests and responses) and a trend, we consider the Hodrick-Prescott (HP) filter \cite{hodrick1997postwar} to be a good candidate. The Hodrick-Prescott filter separates a time-series $y_t$ into a trend $\tau_t$, a cyclical component $\zeta_t$ and an error component $\epsilon_t$ such that:
\[
    y_t=\tau_t+\zeta_t+\epsilon_t
\]
The components are determined by minimising the following quadratic loss function:
\[
\min_{\tau_t}(\sum_{t=1}^{T}(y_t-\tau_t)^2+\lambda\sum_{t=2}^{T-1}[(\tau_{t+1}-\tau_t)-(\tau_t-\tau_{t-1})]^2)
\]
with the multiplier $\lambda$ specifying the penalty offered by the second term. Figure \ref{fig:trend} shows the trends of the different variables representing the bots following the application of the HP filter. In this instance, we picked Alexa, Cryptolocker and Goz. Below each plot, there is also an autocorrelation plot showing the autocorrelation of the respective variable. Unsurprisingly Alexa's autocorrelation (Figure \ref{fig:alexa_trend_auto}) drops quickly as this is not a time series and each observation (domain) is independent of the previous ones. Cryptolocker is quite distinct from Alexa, but Goz shows some similarity.

Autocorrelation is an indication of the number of previous observations (lags) affecting the current observation. It was established that for all bots a minimum of 4 lags is sufficient to reconstruct the whole series. By running an autoregressive moving average (ARMA) estimation over all series, we obtain the coefficients summarised in Table \ref{tbl:arma}.

\begin{table}[th!]
    \centering
    \begin{scriptsize}
    \begin{tabular}{l c c c c c}
    \toprule
      & \textbf{constant} & \textbf{lag(-1)} & \textbf{lag(-2)} & \textbf{lag(-3)} & \textbf{lag(-4)} \\
     \midrule
    \multirow{3}{*}{Alexa} & 3.418966 & 3.088595 & -3.624877 & 1.913501 & -0.383532 \\
                            & \{0.469880\} & \{0.029306\} & \{0.082269\} & \{0.082175\} & \{0.029205\} \\
                            & [0.000] & [0.000]& [0.000]& [0.000]& [0.000]\\
   \midrule
    \multirow{3}{*}{Conficker}   &2.785166 &	3.107565 	&-3.657214 &	1.920612 	& -0.3758890 \\
                &\{4.716197e-01\} &	\{0.029506\} 	& \{0.083259\} &	\{8.340822e-02\} &	\{2.967494e-02\}\\
                 & [0.000] & [0.000]& [0.000]& [0.000]& [0.000]\\
    \midrule
    \multirow{3}{*}{Cryptolocker} & 2.843495 &  	2.880937 & -3.301188 & 1.851547 & 	-4.361981e-01\\
    & \{5.729647e-01\} &\{0.028428\} &\{0.075057\} &\{7.451386e-02\} &\{2.778673e-02\}   \\
                     & [0.000] & [0.000]& [0.000]& [0.000]& [0.000]\\
    \midrule
    \multirow{3}{*}{Goz}   & 3.497740&3.104442&	-3.665087 &1.944414 &-3.897455e-01\\
& \{5.019608e-01\} & \{0.029296\} & \{0.082379\} & \{8.240207e-02\} & \{2.932082e-02\}   \\
                 & [0.000] & [0.000]& [0.000]& [0.000]& [0.000]\\
    \midrule
    \multirow{3}{*}{Matsnu} & 3.111954e+00 &	3.152710 &-3.77899 &2.032400e+00 & -4.112938e-01 \\
 & \{4.716860e-01\} & \{0.028967\} & \{0.08182\} & \{8.174793e-02\} & \{2.888826e-02\}  \\
                 & [0.000] & [0.000]& [0.000]& [0.000]& [0.000]\\    
    \midrule
\multirow{3}{*}{NewGoz}  &2.243034&3.168183&-3.822154&2.070548&-4.203520e-01\\
& \{0.522166\}& \{0.029030\}& \{0.082280\}& \{8.274406e-02\}& \{2.957008e-02\} \\
                 & [0.000] & [0.000]& [0.000]& [0.000]& [0.000]\\
    \midrule
\multirow{3}{*}{Pushdo} & 3.645356e+00&3.110475&-3.685589&1.962390&-3.936961e-01 \\
& \{5.041236e-01\}& \{0.029258\}& \{0.082007\}& \{8.174939e-02\}& \{2.896300e-02\}\\
                 & [0.000] & [0.000]& [0.000]& [0.000]& [0.000]\\
    \midrule
\multirow{3}{*}{Ramdo} & 3.359594&3.164252&-3.842610&2.126389&-4.538455e-01 \\
& \{4.817741e-01\}& \{ 0.028482\}& \{0.080164\}& \{8.033351e-02\}& \{2.865658e-02\}\\
                 & [0.000] & [0.000]& [0.000]& [0.000]& [0.000]\\
    \midrule
\multirow{3}{*}{Rovnix} & 4.380914&3.092777&-3.652384&1.954086&-4.018180e-01 \\
& \{5.473323e-01\}& \{ 0.029101\}& \{0.081594\}& \{8.163828e-02\}& \{2.914092e-02\} \\
                 & [0.000] & [0.000]& [0.000]& [0.000]& [0.000]\\
    \midrule
\multirow{3}{*}{Tinba} & 0.878688&3.380042&-4.351385&2.533331&-5.634898e-01 \\
&\{0.304202\}&\{0.025792\}&\{0.073868\}&\{7.300452e-02\}&\{2.479662e-02\\
                 & [0.000] & [0.000]& [0.000]& [0.000]& [0.000]\\
    \midrule
\multirow{3}{*}{Zeus} &6.110055e+00& 2.773122& -3.025082& 1.637556 & -3.959203e-01\\
& \{7.452222e-01\} & \{0.029171\} & \{0.076807\} & \{7.675001e-02\} & \{2.908530e-02\}\\
                 & [0.000] & [0.000]& [0.000]& [0.000]& [0.000]\\
    \bottomrule
    \end{tabular}
    \end{scriptsize}
    \caption{ARMA estimation for all bots - Std. Error in curly brackets, Probability in square brackets}
    \label{tbl:arma}
\end{table}

To utilise ARMA coefficients as IoCs, we need to investigate whether each tuple can uniquely identify the bot. This is examined by testing whether statistical differences between the different IoCs (coefficients) exist. In addition, we need to test whether subsets of each series can produce the same IoCs as well as the minimum number of (consecutive) observations needed to produce them.

The investigation of the existence of statistical differences was performed as follows. A single variable was produced by stacking all variables and adding a second variable to label the origin of the data and observations. That is, the first 1000 observations were from Alexa (with the label ID=0) the following observations were from Conficker (ID=1), then Cryptolocker (ID=2) and so forth, in accordance with the order displayed in Table \ref{tbl:arma}. We then ran an analysis of variance (ANOVA) with Duncan's test to perform the clustering. The results are shown in Table \ref{tbl:anova}. From the tests, 10 distinct groups emerged, where all bots were successfully separated except Goz and Tinba which were grouped together.

\begin{table}[!ht]
    \centering
    \begin{small}
    \begin{tabular}{lccccc}
    \toprule
         & \textbf{Sum of squares} & \textbf{df} & \textbf{Mean square} & \textbf{F} & \textbf{Sig.} \\
    \midrule
    Between Groups     & 2986937.080 & 10 & 298693.708 & 2310.923 & 0.000 \\
    Within Groups & 1419585.793 & 10983 & 129.253 & & \\
    Total & 4406522.873 & 10993 & & & \\
    \bottomrule
    \end{tabular}
    \end{small}
    
    \caption{ANOVA and clustering results}
    \label{tbl:anova}
\end{table}

Finally, by performing t-tests, we measured that in most cases it requires between 20 to 30 observations to construct the IoCs (ARMA coefficients). From a practical perspective, a network administrator by observing the encrypted network traffic and capturing the lengths of requests and responses, would be able to detect the bot infection, identify the bot; if this is already in the IoC database. Moreover, she may escalate by taking action before the bot completes the communication with the botmaster - that is, if the communication needs more than 30 DNS lookups.

Similar results were obtained with the analysis of the TLS traffic. It is worth mentioning that the distinction between Alexa and non-Alexa (i.e. bot) traffic was stronger, but the distinction between the bots is less apparent. In other words, it is possible to identify the anomalous activity and flag the botnet infection, but attributing to the particular bot would be less effective.


%


Evidently, in both cases, we follow the same steps; therefore, it is needed to investigate the complexity that the proposed traffic analysis implied with regard to the length of exchanged information. While most steps are linear, e.g. collecting the data, applying the Hodrick-Prescott filter etc., the ANOVA method is quadratic. Therefore, the whole process of the proposed traffic analysis has quadratic complexity to the length of exchanged information.

\section{Conclusions}
\label{sec:conclusions}
The continuous arms race between malware authors and anti-malware has made modern malware to use a wide range of cryptographic and obfuscation methods. Therefore, modern malware has become far more stealthy making the quest for novel methods to detect and classify malware a big challenge.

Trying to overcome future challenges, we investigate the use of covert channels for making DNS queries. The motivation towards this direction is that DGAs are detected via the amount and rate of NXDomain responses and the use of pattern matching and machine learning to classify the queries based on their entropy. More recently, in order to circumvent the latter detection, some DGAs have started using combinations of words so that the queries do not appear so random. Nevertheless, the amount and rate of NXDomain responses remains the same. 

We conjunct that in order to circumvent this limitation, malware might soon resort to encrypted DNS queries. As we discussed in our work, this shift would render many of currently deployed security mechanisms useless as they heavily depend on monitoring the unencrypted traffic with the DNS servers. The currently available methods and protocols that are provided by major DNS servers; including Google, Cloudflare, and OpenDNS, if used properly, may allow infected hosts to perform encrypted DNS queries masqueraded as typical HTTPS traffic with whitelisted domains, bypassing this way all deployed security mechanisms. 

To address this threat, we showcase how traffic analysis can be used to provide a lightweight security mechanism and construct IoCs. Using domain names generated from several DGAs we showcase that there are emerging patterns that allow us to identify and classify them accurately. Indeed, despite their covert nature, we illustrate that the Hodrick-Prescott filter can classify them using around 30 samples with very high accuracy and perform bot attribution.

\section*{Acknowledgments}
This work was supported by the European Commission under the Horizon 2020 Programme (H2020), as part of the projects YAKSHA (Grant Agreement no. 780498),  CyberSec4Europe (\url{https://www.cybersec4europe.eu}) (Grant Agreement no. 830929) and the Marie Sk\l{}odowska-Curie grant agreement No 778229 (Ideal-Cities).

The content of this article does not reflect the official opinion of the European Union. Responsibility for the information and views expressed therein lies entirely with the authors.

\section*{References}
\bibliographystyle{elsarticle-num}
\bibliography{refs}
\end{document}